\documentclass[%
 reprint,
nofootinbib,
 amsmath,amssymb,
 aps,
floatfix,
]{revtex4-2}

\usepackage[utf8]{inputenc}

\usepackage{graphicx}
\usepackage{dcolumn}
\usepackage{bm}
\usepackage{hyperref}

\newcommand{\obs}{\mathcal{O}}
\newcommand*{\hk}[0]{\mathbf{h}_{i, x}^k}
\newcommand*{\wk}[0]{\mathbf{w}_{i, x}^k}

\newcommand*{\dk}[0]{\mathbf{d}_{i, x}^k}
\newcommand*{\hl}[0]{\mathbf{h}_{i, x}^\ell}
\newcommand*{\wl}[0]{\mathbf{w}_{i, x}^\ell}
\newcommand*{\ssl}[0]{\mathbf{s}_{i, x}^\ell}
\newcommand*{\dl}[0]{\mathbf{d}_{i, x}^\ell}
\newcommand*{\dlm}[0]{\mathbf{d}_{i, x}^{\ell-1}}
\newcommand*{\valpha}[0]{\alpha_{i, x}}

\DeclareMathOperator{\Real}{Re}

\DeclareMathOperator{\sgn}{sgn}
\DeclareMathOperator{\var}{var}
\DeclareMathOperator{\E}{\mathbb{E}}
\DeclareMathOperator*{\argmin}{arg\,min}
\DeclareMathOperator*{\arcosh}{arcosh}

\usepackage{url}
\hypersetup{
colorlinks=true,
linkcolor=purple,
urlcolor=magenta,
citecolor=blue
}
\usepackage{microtype}
\usepackage{enumitem} 

\usepackage{tcolorbox}
\usepackage[framemethod=TikZ]{mdframed}
\newenvironment{algorithm}[1][]{%
\mdfsetup{%
frametitle={%
\tikz[baseline=(current bounding box.east),outer sep=0pt]
\node[anchor=east,rectangle,fill=black!20]
{\strut \sffamily #1};}}%
\mdfsetup{innertopmargin=10pt,linecolor=black!20,%
linewidth=2pt,topline=true,%
frametitleaboveskip=\dimexpr-\ht\strutbox\relax
}
\vspace{20pt}%
\begin{mdframed}[nobreak=true]\relax%
}{\end{mdframed}}

\graphicspath{ {./img/} }

\begin{document}


\title{Efficient Modelling of Trivializing Maps for Lattice $\phi^4$ Theory Using Normalizing Flows: A First Look at Scalability}

\author{Luigi Del Debbio}%
\author{Joe Marsh Rossney}
 \email[Author to whom correspondence should be addressed --- ]{Joe.Marsh-Rossney@ed.ac.uk}
\author{Michael Wilson}

\affiliation{%
Higgs Centre for Theoretical Physics,
School of Physics and Astronomy,
The University of Edinburgh,
Edinburgh EH9 3FD, UK
}%

\begin{abstract}
General-purpose Markov Chain Monte Carlo sampling algorithms suffer from a dramatic reduction in efficiency as the system being studied is driven towards a critical point through, for example, taking the continuum limit. 
Recently, a series of seminal studies suggested that \textit{normalizing flows} --- a class of deep generative models --- can form the basis of a sampling strategy that does not suffer from this `critical slowing down'. 
The central idea is to use machine learning techniques to build (approximate) trivializing maps, i.e. field transformations that map the theory of interest into a `simpler' theory in which the degrees of freedom decouple. 
These trivializing maps provide a representation of the theory in which all its non-trivial aspects are encoded within an invertible transformation to a set of field variables whose statistical weight in the path integral is given by a distribution from which sampling is easy. 
No separate process is required to generate training data for such models, and convergence to the desired distribution is guaranteed through a reweighting procedure such as a Metropolis test.
From a theoretical perspective, this approach has the potential to become more efficient than traditional sampling since the statistical efficiency of the sampling algorithm is decoupled from the correlation length of the system.
The caveat to all of this is that, somehow, the costs associated with the highly non-trivial task of sampling from the path integral of an interacting field theory are transferred to the training of a model to perform this transformation. 

In a proof-of-principle demonstration on two-dimensional $\phi^4$ theory, \citeauthor{Albergo2019}~\cite{Albergo2019} modelled the trivializing map as a sequence of pointwise affine transformations. 
We pick up this thread, with the aim of quantifying how well we can expect this approach to scale as we increase the number of degrees of freedom in the system.
We make several modifications to the original design that allow our models learn more efficient representations of trivializing maps using much smaller neural networks, which leads to a large reduction in the computational cost required to train models of equivalent quality.
After making these changes, we find that sampling efficiency is almost entirely dictated by how extensively a model has been trained, while being unresponsive to further alterations that increase model flexibility.
However, as we move towards the continuum limit the training costs scale extremely quickly, which urgently requires further work to fully understand and mitigate.
\end{abstract}

\maketitle


\section{Introduction}

Lattice field theory involves the computation of expectation values of the form
\begin{equation} \label{eq:expec}
    \langle \mathcal{O} \rangle
    = \frac{1}{\mathcal{Z}} \int \mathcal{D} \phi \,
    e^{-S(\phi)}    
    \mathcal{O}(\phi) \, ,
\end{equation}
where
\begin{equation}
    \mathcal{D} \phi = \prod_{x\in\Lambda} d\phi_x \, ,
\end{equation}
resulting from the discretisation of Euclidean path integrals onto a space-time lattice $\Lambda$.
$\mathcal{O}(\phi)$ is a generic observable defined for the field configuration $\phi$, and the (Euclidean) action $S(\phi)$ encodes all of the dynamics and interactions of the fields.

In most interesting scenarios integrals of this form are not tractable and we are forced to resort to sampling.
Concretely, estimating \eqref{eq:expec} by sampling means evaluating a statistical average,
\begin{equation} \label{eq:estimator}
    \bar{\obs}
    = \frac{1}{N} \sum_{\{\phi\}} 
    \mathcal{O}(\phi) \, ,
\end{equation}
over a representative sample $\{\phi\}$ comprising $N$ field configurations drawn from a statistical ensemble with Boltzmann factor $e^{-S(\phi)}$.
The error on this estimator scales as $1/\sqrt{N_\text{eff}}$, where the effective sample size $N_\text{eff}$ reaches a maximum value of $N$ in the absence of correlations between configurations.

Markov Chain Monte Carlo (MCMC) methods are the best known tool for sampling from high-dimensional distributions.
However, the configurations in the resulting sequence are indeed correlated, and the effective sample size is diminished by a factor of twice the \textit{integrated autocorrelation time},\footnote{
   The variance of $\bar{\obs}$ is $(2\tau_\obs/N)\Gamma_\obs(0)$, with $N / (2\tau_\obs)$ taken to be the definition of $N_\text{eff}$.
}
\begin{equation} \label{eq:tau_int}
    \tau_\obs = \frac{1}{2} + \sum_{t=1}^\infty \frac{\Gamma_\obs(t)}{\Gamma_\obs(0)} \, ,
\end{equation}
defined for each observable in terms of its autocorrelation function $\Gamma_\obs(t) = \langle \obs(\phi^{(n+t)}) \obs(\phi^{(n)}) \rangle - \langle \obs \rangle^2$, where $t$ represents a number of steps separating pairs of configurations in a Markov chain, and $n$ is arbitrary provided the process has equilibrated to its stationary distribution \cite{Sokal1997}.

Under normal conditions this issue is manageable.
Most MCMC algorithms, however, suffer from an acute condition known as \textit{critical slowing down} associated with a quite catastrophic reduction in their sampling efficiency as the system under study approaches a critical point \cite{Wolff1990}.
Critical slowing down typically manifests as a power-law scaling of the integrated autocorrelation time with the system's correlation length $\xi$,
\begin{equation} \label{eq:csd}
    \tau_\obs \propto \xi^{z_\obs} \, .
\end{equation}
This is rather unfortunate since $\xi$ (in lattice units) diverges as we take the continuum limit of our lattice field theory.

Algorithms based on random-walks or classical molecular dynamics \cite{Metropolis1953,Hastings1970,Duane1987} have lower limit of $z_\obs=1$ owing to the maximum speed of information propagation, but in the absence of very careful tuning \cite{Kennedy1991} they typically exhibit $z_\obs=2$ scaling, corresponding to diffusive information transport.
Furthermore, there is substantial evidence that the picture is even worse when considering theories which possess non-trivial topology in the continuum limit \cite{Campostrini1992,DelDebbio2002,DelDebbio2004,Flynn2015,Bonati2018a}, including QCD itself \cite{Alles1996,Schaefer2011}.
As the continuum limit is approached, the rapid increase in energy barriers between topological sectors can result in $z_\obs > 2$ for topological observables, and potentially even exponential scaling \cite{Vicari1993}.

Collective update algorithms can, in principle, fare much better, since they need not be restricted to local dynamics.
Indeed, several celebrated algorithms based on collective updates have been devised for certain systems \cite{Swendsen1987,Wolff1989,Kusnezov1993,Evertz1993,Prokofev1998}, but as yet a general-purpose collective update algorithm is lacking, and critical slowing down remains an unsolved problem in the majority of cases, including lattice QCD.

Given this state of affairs, there has recently been a great deal of interest in augmenting the MCMC `toolkit' with novel techniques from the rapidly maturing field of machine learning.
Perhaps the most eye-catching new additions to the toolkit are \textit{deep generative models} \cite{Xu2014,BondTaylor2021}, which can be viewed as a category of highly flexible statistical models that, using stochastic optimisation techniques (a.k.a `training'), can approximate complicated probability densities.
The unique property of generative models is that, once trained, they can be directly sampled from to generate samples of potentially `realistic' data.
The relevance of this becomes clear upon a change of terminology: let `data' mean `field configurations' and `probability density' refer to their statistical weight in the path integral.
Already, a number of prototypical hybrid algorithms have been proposed in which deep generative models either guide or replace traditional MCMC update procedures \cite{Torlai2016,Wang2017,Huang2017,Tanaka2017,Liu2017b,Morningstar2017,Urban2018,Singh2020,Albergo2019,Kanwar2020,Nicoli2020,Boyda2020,Albergo2021,Lawrence2021,Foreman2021,Wu2021}.
Deep generative models have also been used to automatically identify relevant variables, leading to novel algorithms for characterising phase diagrams \cite{Cristoforetti2017,Zhou2019,Wang2020,Singh2020,Bachtis2020} and enacting renormalization group transformations \cite{Mehta2014,KochJanusz2018,Lenggenhager2018,Li2018,Efthymiou2019}.

\textit{Normalizing flows} \cite{Tabak2010,Tabak2012,Rezende2015} are a class of deep generative model which approximate the distribution of interest by learning an invertible map from a set of `latent' variables whose distribution is much easier to sample from.
Typically, the map is built out of a sequence of relatively simple pointwise transformations.
The capacity to model complex, correlated probability distributions such as those corresponding to Euclidean lattice field theories arises due to the fact that the parameters of these transformations are generated by neural networks that take the field variables themselves as inputs \cite{Dinh2014,Dinh2016}.
Put another way, the process of training such a model is an encoding of the correlations between the degrees of freedom in the path integral into the weights and biases of these neural networks.

In spirit, normalizing flows are very similar to L\"{u}scher's \textit{trivializing maps} \cite{Luscher2009}, which are field transformations that map an interacting theory to a limit where the field variables decouple, at which point sampling becomes extremely efficient.
Reference~\cite{Luscher2009} provides a power-series expression for the generators of a class of flows which trivialize lattice gauge theories, although only the first two terms in this series are tractable in practice, and additional finite-step errors are accumulated through numerical integration of the flow.
Unfortunately, the degree to which the result of this procedure approximates a trivializing map proved insufficient to improve the scaling of integrated autocorrelation times~\cite{Engel2011}.
However, there has recently been a renewed interest in the potential to construct (approximate) trivializing maps with additional leverage provided by modern machine learning techniques~\cite{Albergo2019,Kanwar2020,Nicoli2020,Boyda2020,Albergo2021,Lawrence2021}.

As noted in Reference~\cite{Luscher2009}, the existence of a trivializing map\footnote{Proof of existence is stated in Reference~\cite{Luscher2009} for the case of compact, connected gauge fields. See Reference~\cite{Luscher2015} for an alternative, diagrammatic construction of trivializing maps which holds to all orders for scalar theories.} implies that it is theoretically possible to evaluate Equation~\eqref{eq:expec} without MCMC, by simply generating uncorrelated random fields.
In Sections \ref{sec:sampling:generative} and \ref{sec:flows} we outline a more practical procedure in which a normalizing flow generates statistically independent field configurations that act as proposals for the Metropolis-Hastings algorithm.
From a theoretical perspective, this has the potential to become more efficient than traditional sampling, since the statistical efficiency of the sampling algorithm is decoupled from the correlation length of the system.
The caveat is that, somehow, the costs associated with the highly non-trivial task of sampling from the path integral of an interacting field theory are transferred to the training of the model.
Therefore, to answer the question of whether a generative sampling algorithm can be expected to outperform traditional methods is a matter of understanding how these training costs scale as the continuum limit is approached.

\citeauthor{Albergo2019} \cite{Albergo2019} first demonstrated that the procedure just described is a viable approach to sampling in lattice field theory.
In their proof of principle study, which focused on two-dimensional scalar $\phi^4$ theory on lattices with up to $14^2$ sites, the normalizing flow was a sequence of pointwise affine transformations parameterised by neural networks.
Here, we continue the same thread, with the aim of establishing how well this approach scales to lattices with up to $20^2$ sites.
Finding that the recipe used in Reference \cite{Albergo2019} yields relatively inefficient representations of trivializing maps for the particular theory of interest, we make a number of adjustments; most importantly, we bring in a more expressive transformation based on a \textit{spline}, and replace the deep neural networks by networks with a single hidden layer that is rather narrow.
We quantify the performance and scaling of our models using hardware-independent metrics: the Metropolis-Hastings acceptance rate, the number of trainable parameters in the models, and the total number of field configurations generated during the training phase.


\section{Sampling in lattice field theory} \label{sec:sampling}

The problem we are trying to solve can be phrased as follows: we would like to generate samples $\{\phi^{(1)}, \phi^{(2)}, \ldots, \phi^{(N)}\}$ of the discrete random field $\{\phi_x \mid x\in\Lambda\} \equiv \phi \in \mathcal{M}^{|\Lambda|}$, where $|\Lambda|$ is the number of sites on the lattice and $\mathcal{M}^{|\Lambda|}$ is a direct product called the field space, that are representative of the lattice field theory we seek to study.
By `representative' we mean that the probability of a particular configuration appearing in the sample is to be proportional to its Boltzmann weight,
\begin{equation} \label{eq:target_density}
    p(\phi) \equiv \frac{e^{-S(\phi)}}{\mathcal{Z}} \, .
\end{equation}
We will refer to $p(\phi)$ as the \textit{target density}.

\subsection{Markov Chain Monte Carlo} \label{sec:sampling:mcmc}

MCMC sampling methods work by generating a sequence of transitions $\phi^{(n)} \to \phi^{(n+1)}$ which together comprise a Markov chain $(\phi^{(n)})^N_{n=1}$.
Thus, implicit in any MCMC method is a transition kernel $W(\phi\to\phi')$, which is required to have a stationary distribution that is equal to the distribution from which we wish to sample, implying the following:
\begin{equation}
    \label{eq:stationarity}
    \int \mathcal D\phi\, p(\phi) W(\phi \to \phi') = p(\phi') \, .
\end{equation}
If $W(\phi\to\phi')$ is also ergodic, then the stationary distribution is unique and the Markov chain is guaranteed to converge to $p(\phi)$ \cite{Chung1967}.
However, this does not imply that any finite section of the chain is representative of $p(\phi)$, since the configurations will be correlated.
In practice, this results in statistical errors that scale as $(2\tau_\obs N)^{-1/2}$ rather than $N^{-1/2}$, leading to a trade-off between algorithmic efficiency --- the amount of effort taken to generate a transition --- and statistical efficiency --- how many transitions are required to produce a statistically independent configuration.

To guarantee Equation~\eqref{eq:stationarity} it is sufficient to impose detailed balance,
\begin{equation}\label{eq:detailed_balance}
    p(\phi) W(\phi\to\phi') = p(\phi') W(\phi'\to\phi) \, .
\end{equation}
Generating transitions that satisfy Equation~\eqref{eq:detailed_balance} is fairly straightforward when the new state differs from the old at only one lattice site, since they amount to sampling from low-dimensional distributions conditioned on the current state of the rest of the lattice.
Collective updates are a different matter entirely; the famous examples \cite{Swendsen1987,Wolff1989,Kusnezov1993,Evertz1993,Prokofev1998} involve contrived update procedures which are only applicable within certain classes of models.

A simple and robust alternative due to \citeauthor{Metropolis1953} \cite{Metropolis1953,Hastings1970} is to generate configurations $\phi'$ via a distribution $q(\phi' \mid \phi)$ which is easy to sample from, and accept or reject these proposals based on an acceptance probability $A(\phi\to\phi')$ such that $W(\phi\to\phi') = q(\phi' \mid \phi) A(\phi\to\phi')$ and detailed balance is satisfied.
The standard choice is the `Metropolis test',
\begin{equation} \label{eq:mh_accept}
    A(\phi \to \phi') = \min \bigg( 1, \, \frac{q(\phi \mid \phi')}{q(\phi' \mid \phi)}
    \frac{p(\phi')}{p(\phi)} \bigg) \, ,
\end{equation}
which, importantly, does not require the calculation of normalizing factors.
A rejection of the proposal corresponds to a duplication of the current state in the chain.
Thus, the Markov chain can be seen as a reweighting of the set of proposals in which the configurations pick up integer weights.
The proposal distribution $q(\phi' \mid \phi)$ can be anything which guarantees ergodicity of $W(\phi\to\phi')$, and it is sufficient for it to have non-zero density everywhere on $\mathcal{M}^{|\Lambda|}$ \cite{Tierney1994}.

In the following snippet of Python code, which implements the Metropolis-Hastings algorithm, \texttt{generator} yields proposals drawn from $q(\phi'\mid\phi)$, and \texttt{acceptance} is a function which evaluates Equation~\eqref{eq:mh_accept}.
\vspace{-.5cm}
\begin{algorithm}[Metropolis-Hastings Algorithm (Python)]
    \ttfamily
    \vspace{-.2cm}
    chain = [] \\
    current = next(generator) \# initialise \\
    for n in range(N):\\
    \hphantom{hhh} proposal = next(generator) \\
    \hphantom{hhh} prob = acceptance(current, proposal) \\
    \hphantom{hhh} if rand() < prob: \\
    \hphantom{hhhhhhh} chain.append(proposal) \\
    \hphantom{hhhhhhh} current = proposal \\
    \hphantom{hhh} else: \\
    \hphantom{hhhhhhh} chain.append(current)
\end{algorithm}
The Metropolis-Hastings algorithm is completely agnostic towards the process through which proposals are generated --- be it a single spin flip, a molecular dynamics trajectory or an independent configuration explicitly drawn from some proposal distribution --- provided any `selection bias' is properly accounted for by the factor $q(\phi \mid \phi') / q(\phi' \mid \phi)$.
This makes it very appealing as a kernel around which to construct collective updates algorithms.

The difficulties arise due to the rapidly increasing sparsity of $p(\phi)$ as the number of degrees of freedom increases and as we move towards the continuum, which puts extremely stringent constraints on how proposals may be generated if we are to sample the path integral in an acceptable amount of time.
Two main approaches to this problem are:
\begin{itemize}
    \item \textit{Local updates}: Generate proposals that are close to the current configuration by updating individual lattice sites.
    Changes in $p(\phi)$ can be made arbitrarily small by tuning the step size so as to yield a desired acceptance rate.
    \item \textit{Hybrid Monte Carlo}: Generate proposals by numerically integrating a fictitious Hamiltonian system, and by doing so update all of the lattice sites. In this case it is the number of integration steps that must be balanced against the acceptance rate.
\end{itemize}
Both of these methods become less efficient when we take the continuum limit, as we are forced to trade down on step size to keep the acceptance rate reasonably high, meaning each statistically independent configuration requires more steps to produce.
This is what is meant by critical slowing down, and the decline in statistical efficiency is quantified by the dynamical critical exponent in Equation~\eqref{eq:csd}.

\subsection{A generative approach to global updates} \label{sec:sampling:generative}

Though it might seem extremely ambitious, the simplest possible generative sampling algorithm would be one in which a deep generative model generates entire, statistically independent field configurations with a probability close to their true weight in the path integral.
Let us entertain this ambition.
For reasons which will shortly become clear, we will focus the following discussion on parametric models with an explicit probability density, $\tilde{p}(\phi)$.
The intention is to construct a model and identify a set of model parameters, $\theta$, such that the approximation $\tilde{p}(\phi) \approx p(\phi)$ is a `good' one.
The notion of a `good approximation' is made quantitative through the Kullbach-Leibler divergence \cite{Kullbach1951},
\begin{equation} \label{eq:kl}
    D_\text{KL}(p \; || \; \tilde{p}) = 
    \int \mathcal D\phi \, p(\phi) 
    \log \frac{p(\phi)}{\tilde{p}(\phi)} \, .
\end{equation}
When we speak of `training' such a model, what we really mean is optimising a particular function with respect to the model's parameters.
In the present work, this `objective function' will be (a variant of) the Kullbach-Leibler divergence, and the goal of training will be to find the set of parameters $\theta^*$ which satisfy
\begin{equation} \label{eq:optim}
\theta^* = \argmin_\theta D_\text{KL}(p \; || \; \tilde{p}) .
\end{equation}

Successfully training a generative model to generate uncorrelated samples of field configurations with a probability close to their true weight in the path integral certainly appears to be an auspicious starting point for constructing an efficient sampling algorithm.
Of course, the problem is that anything less than a perfect fit, i.e. $\tilde{p}(\phi) = p(\phi)$, implies that the samples generated by the model are not truly representative of the field theory, with discrepancies between $\tilde{p}(\phi)$ and $p(\phi)$ manifesting as biases in expectation values.
Yet it is possible to exactly correct for these biases through reweighting or a Metropolis step, provided we have access to $\tilde{p}(\phi)$.
Hence, as well as restricting ourselves to models with an \textit{explicit} density function, we will also demand that $\tilde{p}(\phi)$ is \textit{tractable}, by which we mean it is given exactly by a closed-from expression computable in polynomial time (in order to be scalable) and whose repeated evaluation (for us, $10^5-10^9$ times during training) does not constitute an unacceptably large overhead.\footnote{
One might think that likelihood-based training (i.e. solving \eqref{eq:optim}) \textit{requires} $\tilde{p}(\phi)$ to be tractable, but approximate training schemes based on a variational upper bound of \eqref{eq:kl} have proved successful.}

Consider a variant of the Metropolis-Hastings algorithm in which \texttt{generator} is a generative model equipped with an explicit and tractable density that is capable of generating independent configurations with probability given by
\begin{equation}
    \label{eq:ProposalDGM}
    q(\phi' \mid \phi) = \tilde{p}(\phi')\, .
\end{equation}
If the model were a perfect approximation, such that $\tilde{p}(\phi) = p(\phi)$ for all $\phi$, then $A(\phi\to\phi') = 1$ identically and 100\% of proposals would be accepted.
In a more realistic situation where there are discrepancies, the inefficiency of generating proposals with a probability proportional to $\tilde{p}(\phi)$ rather than $p(\phi)$ manifests itself through multiplicities in the Markov chain due to rejections, which are in turn measurable as autocorrelations.
However, if proposals are drawn independently, then rejections are the \textit{only} source of autocorrelation.
As explained in Reference \cite{Albergo2019}, the autocorrelation at separation $t$ is given, \textit{for all observables}, by
\begin{align} \label{eq:tau_rej}
    \frac{\Gamma_\obs(t)}{\Gamma_\obs(0)} &= \Pr ( t \text{ consecutive rejections} ) \nonumber\\
    &= \E_{\phi \sim p} \left[ \left( 
    \E_{\phi' \sim \tilde{p}} \left[
    1 - A(\phi \to \phi')
    \right] \right)^t \right] \, .
\end{align}
This is an extremely appealing feature that is not present in traditional algorithms, where local dynamics combined with energy barriers can lead to the decoupling of autocorrelation times for topological and non-topological observables \cite{DelDebbio2004}.
An estimate of Equation~\eqref{eq:tau_rej} is trivial to obtain from the accept/reject history of a Metropolis-Hastings simulation, with which we can compute an estimate for the integrated autocorrelation time that we denote $\tau_\text{rej}$.

Since Equation~\eqref{eq:tau_rej} is strictly larger than the average rejection rate $\displaystyle 1 - \E_{\phi\sim p}\E_{\phi' \sim \tilde{p}} \left[ A(\phi \to \phi') \right]$ raised to the $t$-th power, a lower bound on the integrated autocorrelation time can be given in closed form by a geometric series,
\begin{equation} \label{eq:lower_bound}
    \tau_\obs \geq 
    \frac{1}{\E_{\phi\sim p} \E_{\phi' \sim \tilde{p}}
    \big[ A(\phi \to \phi') \big]} - \frac{1}{2} \, .
\end{equation}
This expression is not particularly useful per se, but we will be interested in how close to this lower bound the actual integrated autocorrelation falls.

Although this is not the approach we will take, reweighting can instead be done at the level of computing ensemble averages through a change of measure in Equation~\eqref{eq:expec} to $\mathcal{D} \phi \, \tilde{p}(\phi) w(\phi)$, where the reweighting factor $w(\phi) \equiv p(\phi) / \tilde{p}(\phi)$ is the same factor used in the Metropolis test~\cite{Nicoli2020}.
The mean estimator from Equation~\eqref{eq:estimator} then reads
\begin{equation}
\bar{\mathcal{O}} = \frac{\sum_{\phi\in\Phi} w(\phi) \mathcal{O}(\phi)
    }{\sum_{\phi\in\Phi} w(\phi)} \, .
\end{equation}
Of course, while this approach makes use of all of the generated configurations, there is still a price to be paid for drawing samples from $\tilde{p}(\phi)$ rather than $p(\phi)$; the weights ensure that the number of configurations yielding non-negligible contributions to the sum drops rapidly as the approximation $\tilde{p}(\phi) \approx p(\phi)$ degrades.
As remarked on in Reference~\cite{Boyda2020}, this a posteriori reweighting approach is appealing if $\mathcal{O}(\phi)$ is cheap to compute relative to the cost of generating configurations from the model.


\section{Normalizing flows} \label{sec:flows}

For our purposes, we define a normalizing flow as a both-directions continuously differentiable bijective mapping,\footnote{This is precisely the definition of a $C^1$-diffeomorphism.}
\begin{align*}
    f_\theta : \mathcal{M}^{|\Lambda|} &\to \mathcal{M}^{|\Lambda|} \\
    z &\mapsto \phi=f_\theta(z) \, ,
\end{align*}
between `latent' random variables, $z \sim r(z)$, and `candidate\footnote{We refer to the configurations generated by the model are referred to as `candidate' since they might be rejected by the Metropolis test.} field configurations', $\phi = f_\theta(z) \sim \tilde{p}(\phi)$.

We will immediately restrict ourselves to the special case of $\mathcal{M} = \mathbb{R}$, which applies to scalar $\phi^4$ theory.\footnote{For discussion and examples of flows on non-Euclidean manifolds, see References \cite{Gemici2016,Rezende2020}.}
Hence, the density associated with the candidate field configurations is given by the familiar formula for a change of variables, involving the Jacobian determinant,

\begin{align} \label{eq:ch_var_formula}
    \tilde{p}\big(f_\theta(z)\big) = r(z) \bigg| \frac{\partial f_\theta(z)}{\partial z} \bigg|^{-1} \, .
\end{align}

In practice, we only every require the logarithm of this equation.

We will draw latent variables from an uncorrelated Gaussian distribution,

\begin{equation} \label{eq:prior}
    r(z) = \prod_{x\in\Lambda} \frac{1}{\sqrt{2\pi\sigma^2}} e^{-z_x^2 / (2\sigma^2)} \, ,
\end{equation}

which one may interpret, in the spirit of References \cite{Luscher2009,Engel2011}, as a trivial limit of $\phi^4$ theory.\footnote{Cf. Equation~\eqref{eq:phi_four:action} with $\lambda\to 0$ and $\beta\to 0$.}
In principle one could put more effort into generating latent variables that reduce the workload for the flow.\footnote{In Section \ref{sec:outlook:scalability_physics} we briefly discuss the use of free fields as the latent variables.}
However, even with a priori knowledge of some features of the target density this comes with a high risk of over-engineering the problem; i.e. leading to marginal improvements in the model's approximation to the target, which are completely negated by the increased costs of generating samples and computing the density.

Although the bijective construction is not the most flexible a priori, normalizing flows have several advantages over other generative models.
Crucially, it is straightforward in principle to ensure that the density is tractable, by choosing a map whose Jacobian determinant $\lvert \partial f_\theta(z) / \partial z \rvert$ is tractable.
It is this feature that provides us with a means of guaranteeing convergence to the correct target density through the Metropolis test.
Additional benefits relate to the training, which is discussed in the next subsection.
Finally, as an added bonus, the intermediate states of the flow correspond to valid probability densities in their own right, from which we can draw samples.
In this sense, Normalizing flows are more `interpretable' than other generative models which can behave more like a `black box'.

\subsection{Training a flow model} \label{sec:flows:training}

Since normalizing flows are differentiable by construction, they can be trained using standard gradient-based optimisation algorithms such as stochastic gradient descent.
The algorithm used in this work is a variant that incorporates momentum, called ADAM~\cite{Kingma2014}.
The conventional approach to training is to expose the model to a set of data drawn from the distribution of interest, $p(\phi)$, via a separate process and tune the parameters of the model in order to optimise some objective function.
If we were to take the conventional approach here, the set of training configurations would be divided into `batches', passed through the layers of the flow model in the reverse direction, and the resulting variables $f_\theta^{-1}(\phi)$ used to estimate the following objective function by averaging over the batch:
\begin{widetext}
\begin{equation} \label{eq:kl_estimator}
    \hat{D}_\text{KL}(p \; || \; \tilde{p}) =
    \E_{\phi \sim p(\phi)}
    \bigg[ 
    - \log r\big( f_\theta^{-1}(\phi) \big)
    - \log \bigg| \frac{\partial f_\theta^{-1}(\phi)}{\partial \phi} \bigg|
    \bigg] + \text{irrelevant terms}\, .
\end{equation}
\end{widetext}
Equation~\eqref{eq:kl_estimator} is an estimator for the Kullbach-Leibler divergence defined in Equation~\eqref{eq:kl}, up to an unknown self-information term, $\E_{\phi\sim p(\phi)}[\log p(\phi)]$, that does not depend on the model's parameters and is therefore irrelevant for the purposes of optimisation.

However, for our purposes this strategy is clearly not satisfactory since the problem has gone full circle; the ability to train models would then be tied to the ability to generate a large representative samples of configurations to act as training data, which is exactly what we are prevented from doing by critical slowing down.
Thankfully, an alternative path presents itself in the typical scenario where we are interested in sampling from a theory for which $S(\phi)$ is completely specified.
In this training paradigm, favoured by Reference~\cite{Albergo2019} and many subsequent studies, one considers the alternative definition of Kullbach-Leibler divergence following a reversal of the arguments with respect to Equation~\eqref{eq:kl},
\begin{equation} 
    \label{eq:kl_reverse}
    D_\text{KL}(\tilde{p}\; || \; p) = \int 
    \mathcal D\phi\,
    \tilde{p}(\phi) 
    \log \frac{\tilde{p}(\phi)}{p(\phi)} \, .
\end{equation}
This allows us to define an objective function that can be minimised using estimates based on configurations generated exclusively by the model:
\begin{widetext}
\begin{equation} \label{eq:kl_estimator_reverse}
    \hat{D}_\text{KL}(\tilde{p} \; || \; p) =
    \E_{z \sim r(z)}
    \bigg[ S\big( f_\theta(z) \big)
    - \log \bigg| \frac{\partial f_\theta(z)}{\partial z} \bigg|
    \,\bigg] + \text{irrelevant terms}\, .
\end{equation}
\end{widetext}
In Equation \eqref{eq:kl_estimator_reverse}, the irrelevant terms that do not depend on the model's parameters are $\E_{z \sim r(z)} [\log r(z)]$ and the normalizing factor in the path integral, $\log \mathcal{Z}$.

One could be forgiven for thinking that the difference between optimising $\hat{D}_\text{KL}(p \; || \; \tilde{p})$ and optimising $\hat{D}_\text{KL}(\tilde{p} \; || \; p)$ is no more than a matter of exchanging an pre-existing training set for configurations drawn from the model.
In practice, however, the two modes of training have distinct quirks which are important to appreciate.
A recent contribution, Reference~\cite{Hackett2021}, includes a comparison of the two training schemes in situations where $p(\phi)$ is multi-modal.

In fact, by insisting on not having to obtain training inputs from an external process, we have actually sidestepped several of the major difficulties normally faced during training.
In particular, since each batch of training inputs is stochastically generated on-demand, and never recycled, `over-fitting' of training data is relegated to a non-issue.
Furthermore, we need not be concerned about bias in the training inputs, since they are obtained through exact sampling from the latent distribution.
The problem that we are most likely to encounter is one of insufficient flexibility to resolve all of the features in the target density, leading to a model which `under-fits' the target.
This is an important example of qualitatively different results arising from the choice of training scheme.
Optimising $\hat{D}_\text{KL}(p \; || \; \tilde{p})$ typically results in `smoothed' approximations to the target, whereas the approach we take, optimising $\hat{D}_\text{KL}(\tilde{p} \; || \; p)$ by sampling from the model, has a tendency to fit (not necessarily all of) the modes of the target, and set $\tilde{p}(\phi) \approx 0$ elsewhere.
This behaviour is explained and its implications discussed in Sections~\ref{sec:discussion:training} and \ref{sec:discussion:warning}.

\subsection{Building flexible models} \label{sec:flows:flexible}

When considering potential transformations for the layers $g_i$, there are two conflicting requirements that will need to be met with a potentially very delicate compromise.
Firstly, the flow will need to be highly flexible in order to start with uncorrelated Gaussian variables and distil the complex features of a system near to criticality, which will include non-trivial correlations on multiple scales.
On the other hand, the Jacobian determinant in Equation~\eqref{eq:ch_var_formula} must be tractable since we are still required to evaluate $\tilde{p}(\phi)$ in order to perform the reweighting that guarantees convergence to the correct target.
Furthermore, the speed at which models can be trained and sampled from will depend on the efficiency with which the (logarithm of) the Jacobian determinant can be computed.
This is a significant constraint and one that is very much at odds with the goal of using invertible transformations to map simple densities to complex ones; it is challenging to define sufficiently expressive transformations without rendering the Jacobian term intractable.

The key component that initially enabled normalizing flows to become competitive with more flexible generative models at performing benchmark tasks (such as image synthesis) was a particular type of transformation now widely referred to as a \textit{coupling layer} \cite{Dinh2014,Dinh2016,Kingma2018}.
Coupling layers are essentially a template for building flexible, pointwise, invertible transformations that are guaranteed to have a triangular Jacobian matrix.
One divides the inputs to a coupling layer into two groups, only one of which will actually undergo a non-trivial transformation that is conditioned on information derived from the remaining, non-transformed variables.
Since we are interested in $\phi^4$ theory with a single degree of freedom at each lattice site, this splitting equates to an (arbitrary) partitioning of the lattice into $\Lambda^A$ and $\Lambda^P$, which we refer to as the `active' and `passive' partitions, respectively.
The normalizing flow is then built out of several couplings layers by function composition, $f_\theta \equiv g_I \circ g_{I-1} \circ \ldots \circ g_2 \circ g_1$.
Defining $v_1 \equiv z$ and $v_{I+1} \equiv \phi$ lets us write the action of the $i$-th coupling layer as
\begin{align*}
    g_i : \mathbb{R}^{|\Lambda|} &\to \mathbb{R}^{|\Lambda|} \\
    v_i &\mapsto v_{i+1} = g_i(v_i) \, ,
\end{align*}
where
\begin{align} \label{eq:coupling_layer}
    v_{i+1,x} = \begin{cases}
        v_{i,x} \, , &x\in\Lambda^P_i\\
        C_{i,x}\big(v_{i,x} ; \mathbf{N}_{i,x}(v_i^P) \big) \, , &x\in\Lambda^A_i \, .
    \end{cases}
\end{align}
$v_i^P \in \mathbb{R}^{|\Lambda^P_i|}$ is a vector built out of elements $\{v_{i, x} \, | \, x \in \Lambda^P_i\}$, with $v_i^A \in \mathbb{R}^{|\Lambda^A_i|}$ defined analogously.
The set of functions $C_i$ (which are as yet unspecified) transform the active partition and are conditioned on a set of parameters $\mathbf{N}_i(v_i^P)$ that are themselves functions of the passive variables. In the examples that we will consider here, these parameters are the output layer of one or more fully-connected feed-forward neural networks.\footnote{
There is an important distinction to be made between the neural network outputs, $\mathbf{N}_{i,x}$, that parametrise a function, $C_{i,x}$, and the `model parameters', $\theta$, that are tuned during training, which are the aggregation of all of the parameters (weights and biases) from each neural network, from each coupling layer.
}
Throughout this paper, neural network outputs, exclusively, will be denoted by bold letters, and it will be left to the presence or absence of indices (e.g. $i$ for the layer index, $x$ for the lattice sites) to specify the cardinality of sets (for example, $v_{i, x}$ is a number while $v_i$ is a vector with $\lvert\Lambda\rvert$ components).

The Jacobian for a coupling layer is, in block notation,
\begin{equation} \label{eq:coupling_jacob}
    \begingroup
    \renewcommand*{\arraystretch}{1.5}
    \frac{\partial g_i}{\partial v_i}
    = \begin{pmatrix}
        \mathbb{I} & 0 \\
        \frac{\partial C_i}{\partial v_i^P} & \frac{\partial C_i}{\partial v_i^A}
    \end{pmatrix}
    \endgroup \, ,
\end{equation}
where the lower-right block is understood to be a diagonal matrix whose diagonal elements are $\partial C_{i,x}/\partial v_{i,x}^A$ for each $x\in\Lambda^A_i$.
As promised, this matrix is triangular, so the determinant is simply equal to the product of terms on the leading diagonal.
\begin{equation} \label{eq:coupling_det_jacob}
    \left\lvert \frac{\partial g_i}{\partial v_i} \right\rvert
    = \left\lvert \prod_{x \in \Lambda^A} \frac{\partial C_{i, x}}{\partial v_{i, x}} \right\rvert \, .
\end{equation}
By swapping the active and passive partition after every coupling layer ($\Lambda^{A/P}_{i+1} = \Lambda^{P/A}_i$) and composing at least three layers, we ensure that each lattice site is updated using information from every other one.
Thus, coupling layers allow us to sample from correlated target densities using uncorrelated latent variables \textit{at no additional expense in the computation of the Jacobian determinant} --- the calculation is the same as it would be if we replaced the neural networks $\mathbf{N}_i$ with parameters that had no dependence on $v_i^P$.

Since $\det AB = \det A \det B$, a sequence of $I$ coupling layers induce the following Jacobian determinant, 
\begin{equation} \label{eq:flow_log_det_jacob}
    \log \bigg| \frac{\partial f_\theta}{\partial z} \bigg| 
    = \sum_{i=1}^{I} \log \bigg| \frac{\partial g_i}{\partial v_i} \bigg| \, .
\end{equation}
These terms can be accumulated alongside the transformations of field variables, so that a single pass through all of the coupling layers yields both a set of candidate field configurations and the left hand side of Equation \eqref{eq:flow_log_det_jacob} for each configuration in the batch, ready to evaluate $\log \tilde{p}(\phi)$ (if sampling) or the objective function (if training).

\begin{figure}
        \includegraphics[width=.5\textwidth]{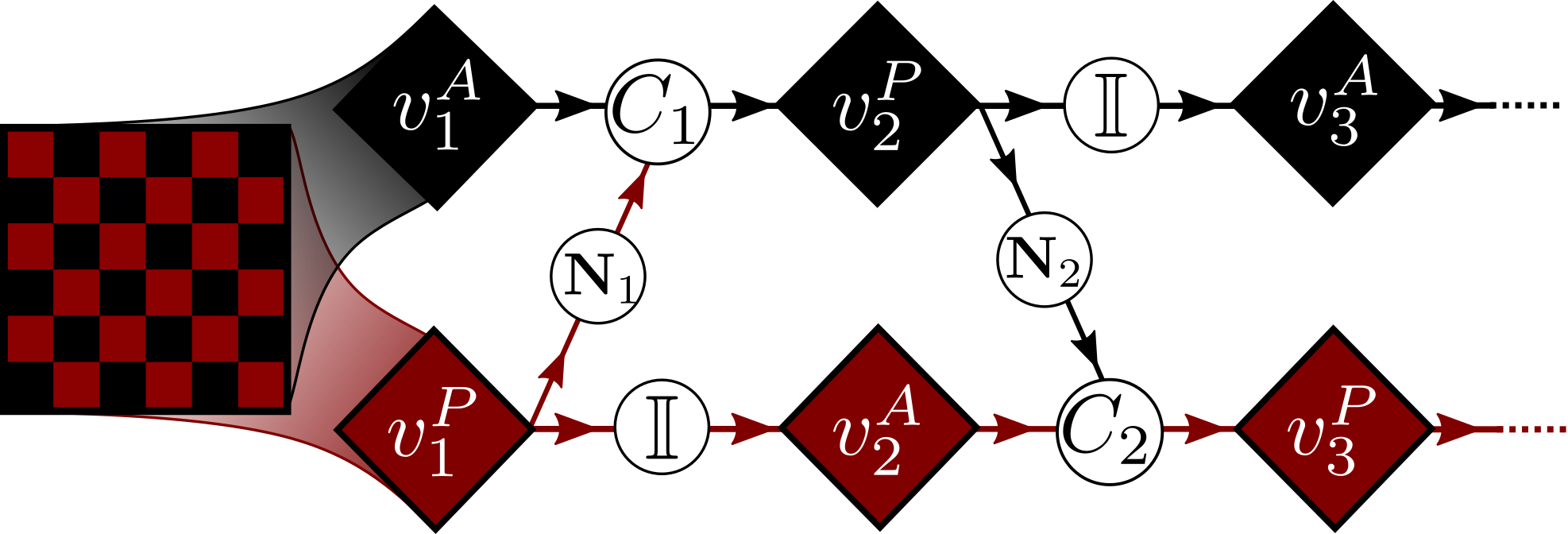}
        \caption{
            Graphical representation of the first coupling block under the `checkerboard' partitioning scheme.
            The latent Gaussian variables, $v_1 \equiv z$, are split into two partitions (the red and black nodes).
            The $i$-th coupling layer transforms the active partition $C_i \; : \; v_i^A \mapsto v_{i+1}^P$ using information from the passive partition, $v_i^P$, via the neural network(s) $\mathbf{N}_i(v_i^P)$.
            $\mathbb{I}$ denotes the identity transformation.
            Note that there is no need to merge the active and passive partitions until after the final coupling layer.
    }
    \label{fig:checkerboard}
\end{figure}

Given freedom to divide the lattice in whichever way seems fit, we will implement a `checkerboard' partitioning featuring in Figure~\ref{fig:checkerboard}, which ensures that each lattice site is directly influenced by its closest neighbours.
We will often refer to a pair of coupling layers, which together transform every degree of freedom once, as a `coupling block'.
From hereon, we will drop the $A$ and $P$ superscripts and assume we are always talking about transforming a set of variables $v_i$ belonging to the active partition.
Furthermore, we will denote the neural networks without an explicit dependence on the passive partition.

\subsection{Affine and additive transformations} \label{sec:flows:affine}

Affine coupling layers were introduced by \citeauthor{Dinh2016} \cite{Dinh2016} as part of the \textit{Real NVP} architecture.
The pointwise transformation multiplies and shifts each degree of freedom, and is commonly written in vector form,

\begin{equation} \label{eq:affine_layer}
    C^\text{aff}_i(v_i \, ; \, \mathbf{s}_i, \mathbf{t}_i) 
    = ( v_i - \mathbf{t}_i) \odot e^{-\mathbf{s}_i}
    \, ,
\end{equation}

where $\mathbf{s}_i$ and $\mathbf{t}_i$ are modelled by neural networks with $|\Lambda^A|$ outputs, and $\odot$ is the element-wise product.

Using Equations~\eqref{eq:coupling_det_jacob} and \eqref{eq:flow_log_det_jacob}, a single affine coupling layer contributes

\begin{equation}
    \label{eq:affine_log_det_jacob}
    \log \bigg| \frac{\partial g_i^\text{aff}}{\partial v_i} \bigg| 
    = -\sum_{x \in \Lambda^A} \mathbf{s}_{i,x} \, 
\end{equation}

to the logarithm of the Jacobian determinant.
The precursor to \textit{Real NVP} uses volume-preserving `additive' coupling layers~\cite{Dinh2014}, such that Equation \eqref{eq:affine_layer} reduces to the shift by $\mathbf{t}_i$ only,

\begin{equation} \label{eq:additive_layer}
    C^\text{add}_i(v_i \, ; \, \mathbf{t}_i) 
    = v_i - \mathbf{t}_i
    \, .
\end{equation}

In our implementation of these coupling layers, we standardise the inputs to the neural networks such that they have unit variance, and do not apply activation functions to the output layer of these neural networks.
We also append a global rescaling transformation after all of the coupling layers have acted, which can have a learnable scale parameter.

As an inexpensive yet remarkably expressive flow architecture, \textit{Real NVP} has achieved widespread success and is frequently taken as a benchmark model to which new flow models are compared.
However, more sophisticated flows using more flexible transformations have since achieved superior results on a number of standard datasets (mostly images) --- see e.g. References~\cite{Papamakarios2017,Kingma2018,Huang2018,Grathwohl2018,Muller2018,Hoogeboom2019,Durkan2019a,Durkan2019b,Meng2020}.
This motivates us to explore one of the prominent alternatives.

\subsection{Rational quadratic splines}

\begin{figure}[b]
        \includegraphics[width=.45\textwidth]{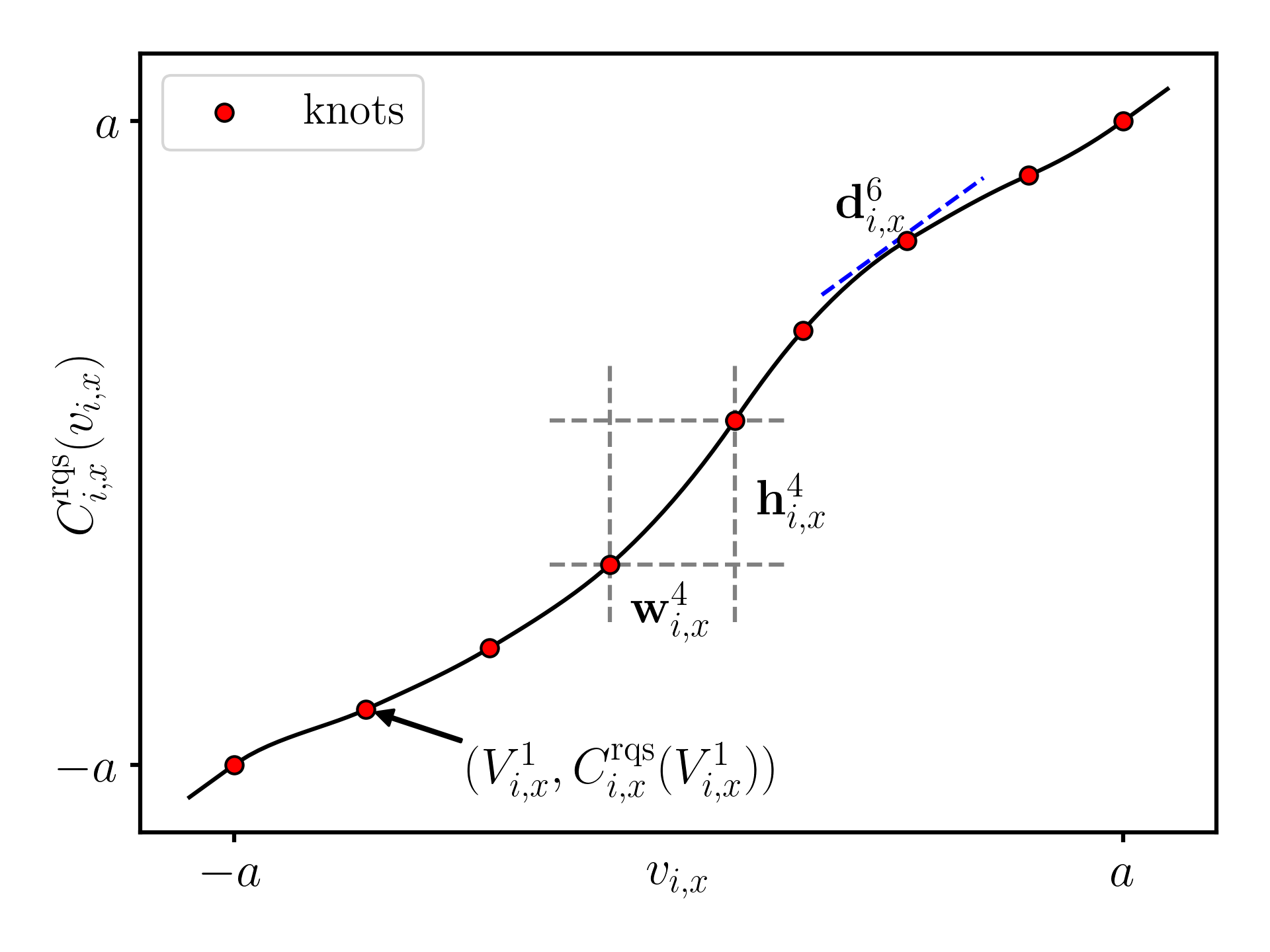}
        \caption{
            An example 8-segment rational quadratic spline transforming the degree of freedom at lattice site $x$.
            $\mathbf{w}^k_{i, x}$ and $\mathbf{h}^k_{i, x}$ are the widths and heights of the rectangle containing the $k$-th polynomial segment.
            $\mathbf{d}_{i, x}^k$ is the derivative at the $k$-th knot.
    }
    \label{fig:example_rqs}
\end{figure}

Splines are functions defined piecewise by polynomials.
Coupling layers using spline-based transformations were introduced in Reference~\cite{Muller2018} and further developed by \citeauthor{Durkan2019b} in References~\cite{Durkan2019a,Durkan2019b}.
We will focus on the most flexible member of the family as described in Reference~\cite{Durkan2019b}, which is based on a continuously differentiable spline interpolant first considered in Reference~\cite{Gregory1983}.

A rational quadratic spline (RQS) transformation $C^\text{rqs}_{i, x}$ is defined for a single degree of freedom by $K$ rational quadratics, referred to as the `segments' of the spline.
These segments are joined end-to-end at a set of `knots' $\big\{\big(V_{i, x}^k, C^\text{rqs}_{i, x}(V_{i, x}^k)\big) \, | \, k = 0, \ldots, K \big\}$ such that the result is a strictly monotonic, $C^1$-differentiable function on the interval $[-a, a]$, which will be chosen in order to contain essentially all of the probability mass.

Given a reference point, the parameterisation provided by Reference~\cite{Gregory1983} requires $3K+1$ strictly positive parameters to uniquely specify this function: the side lengths $(\mathbf{w}^k_{i, x}, \mathbf{h}^k_{i, x})$ of the $K$ rectangles which have adjacent knots on their opposing corners (often referred to as `widths' and `heights'), and the derivatives $\mathbf{d}^k_{i, x}$ at the $K+1$ knots.
Note that our choice of labelling, featuring in Figure \ref{fig:example_rqs}, means that the endpoints of the $k$-th segment are the $(k-1)$-th and $k$-th knots.
This gives us a set of $|\Lambda^A| \times (3K + 1)$ parameters for the coupling layer,

\begin{equation*}
\hspace{-3mm}
    \mathbf{N}_i = \big\{ \wk, \hk, \dk,  \mathbf{d}_{i, x}^0 \, | \, k = 1, \ldots, K ; \; x \in \Lambda^A\big\} .
\end{equation*}

For later convenience, define the slopes of the straight lines connecting adjacent knots as

\begin{equation} \label{eq:rqs_slope}
    \frac{C_{i,x}^\text{rqs}(V_{i,x}^k) - C_{i,x}^\text{rqs}(V_{i,x}^{k-1})}
    {V_{i,x}^k - V_{i,x}^{k-1}}
    = \frac{\mathbf{h}_{i,x}^k}{\mathbf{w}_{i,x}^k} 
    \equiv \mathbf{s}_{i,x}^k
    \geq 0 \, ,
\end{equation}

and re-express the variables being transformed as

\begin{equation} \label{eq:rqs_alpha}
    \frac{v_{i,x} - V_{i,x}^{\ell-1}}
    {\mathbf{w}_{i,x}^\ell}    
    \equiv \alpha_{i,x} 
    \in [0, 1] \, ,
\end{equation}

which corresponds to the fractional position of $v_{i,x}$ within the specific segment in which it is located, whose index we label $\ell$.
Note that each degree of freedom $v_{i, x}$ must first be sorted into the appropriate segment (i.e. the value of $\ell$ determined) using e.g. bisection search, which is not too expensive since the knots are already sorted into ascending order.

Using the 0th knot at $\big( V_{i, x}^0 , C^\text{rqs}_{i, x}(V_{i, x}^0) \big) = ( -a, -a )$ as the reference point, the RQS transformation and its gradient can then be written, for each degree of freedom, using Equations \eqref{eq:rqs_coupling} and \eqref{eq:rqs_grad}.
\begin{widetext}
\begin{equation} \label{eq:rqs_coupling}
    C^\text{rqs}_{i,x} ( v_{i, x} \, ; \, \mathbf{N}_{i, x})
    = \underbrace{-a + \sum_{k=1}^{\ell-1} \hk}_{V_{i,x}^{\ell-1}}
    + \frac{ \hl \big[ \ssl \valpha^2 + \dlm \valpha (1 - \valpha) \big] }
    {\ssl + (\dlm + \dl - 2 \ssl) \valpha(1 - \valpha)} \, ,
\end{equation}
\begin{equation} \label{eq:rqs_grad}
    \frac{1}{\wl} \frac{d C^\text{rqs}_{i, x}} {d\valpha}
    = \frac{ (\ssl)^2 \big[ \dl \valpha^2 + 2\ssl \valpha(1 - \valpha) + \dlm(1 - \valpha)^2 \big]}
    {\big[ \ssl + (\dlm + \dl - 2\ssl) \valpha (1 - \valpha) \big]^2} \, .
\end{equation}
\end{widetext}
Taking the logarithm of Equation~\eqref{eq:rqs_grad} and summing over all $x$ in the active partition yields the contribution to the logarithm of the Jacobian determinant from one RQS coupling layer.

The advantage of using this parameterisation should now be clear; all that is required to guarantee that the gradient is strictly positive is that every parameter in $\mathbf{N}_i$ is also strictly positive.
It is also particularly simple to enforce the desired normalisation:
\begin{equation} \label{eq:rqs_width_norm}
    \sum_{k=1}^K \wk =
    \sum_{k=1}^K \hk = 2a \, . 
\end{equation}
The function defined by Equation~\eqref{eq:rqs_coupling} is an interpolant for the set of knots,\footnote{To see this, substitute $\alpha_{i, x} = 0$ or 1.} so the problem of representing complicated transformations reduces to one of generating a sufficient number of knots with sufficient accuracy.

After fixing $\mathbf{d}_i^0 = \mathbf{d}_i^K = \{1\}^{|\Lambda^A|}$, we let a \textit{single} neural network generate the remaining $|\Lambda^A| \times (3K - 1)$ parameters in the RQS layer, using the $|\Lambda^P|$ field variables in the passive partition as inputs.
We take the unconstrained outputs of the neural net --- denoted below with a hat --- and split them into widths, heights, and derivatives.
Positivity of the $\hk$ and $\wk$, as well as the correct normalisation, is enforced by passing the unconstrained widths and heights through a `softmax' activation function,
\begin{equation}
    \hk = \frac{e^{\widehat{\mathbf{h}}_{i,x}^k}}{\sum_{k'=1}^K e^{\widehat{\mathbf{h}}_{i, x}^{k'}}} \times 2a \, .
\end{equation}
The derivatives are instead passed through a `softplus',
\begin{equation}
    \dk = \log (1 + e^{\widehat{\mathbf{d}}_{i, x}^k} ) \, .
\end{equation}

To ensure that the inputs of a spline layer fall within the interval $[-a, a]$, we generally chose $a = 5$ and standardised the inputs before the first RQS layer by dividing them by the standard deviation, taken over both the batch and the lattice sites.
To catch the edge cases of inputs falling outside of this interval, we extended the definition of the transformation to be the identity outside of $[-a, a]$, while fixing the derivatives at the external knot points to be unity to ensure that the transformation remains everywhere differentiable.

\subsection{Neural networks}

The principle behind this entire approach is that correlations in the target density can be encoded in the weights and biases of neural networks.
Once these weights and biases have been fixed (i.e. once we have finished training), a neural network is simply a function.
In this case, the role of these functions is to take a set of field variables (the passive partition) as inputs, and use this information to return a set of parameters that govern the transformation of a different set of field variables (the active partition).

There are theoretical grounds to believe that this works; specifically, there is a universal approximation theorem implying that any well-behaved function defined on a compact subspace of $\mathbb{R}^{|\Lambda^P|}$ can be approximated with arbitrary accuracy by functions of the form
\begin{equation} \label{eq:mlp}
    \mathbf{N}_x(v^P) = \sum_{j=1}^{H} A^{(2)}_{xj} \, \sigma \Big( \sum_{x'\in\Lambda^P} A^{(1)}_{jx'} v^P_{x'} + b^{(1)}_j \Big)
\end{equation}
which represent feed-forward neural networks with a single hidden layer containing $H$ elements and a non-linear activation function $\sigma$ \cite{Hornik1991}.
$A^{(1)}, A^{(2)}$ are matrices containing the network weights, and $b^{(1)}$ is a vector of biases.

This theorem does not tell us how large the hidden layer ought to be in order to represent the desired function with a given accuracy, though the number will be related to the degree of non-linearity and sensitivity with which the function must respond to variations in its inputs.
Coupling layers will need to be able to dramatically alter the field variables in response to their closest neighbours, while also accounting for more subtle corrections due to field at larger separations, which implies that the functions being modelled by the neural networks must be extremely sensitive.
However, there is no hard-and-fast rule dictating the optimal size and depth of neural networks for our specific problem; this must be discovered through experimentation.

Universal approximation theorems have also been proven, quite recently, for so-called (deep) convolutional neural networks \cite{Yarotsky2018,Zhou2018,Heinecke2020}.
The building blocks of these networks are convolutions, denoted by $\star$, with a set of `kernels', $W$, whose weights are trainable parameters,

\begin{equation} \label{eq:conv}
    W \star v = \sum_{k_1,k_2=0}^{L-1} W(k) v((x-k)_{\mathrm{mod}\,L}) \, .
\end{equation}

In the above, $k = (k_1,k_2)$ and lattice coordinates have temporarily been written as arguments rather than the usual subscripts.
A `stride' size of one is also implied, meaning that the convolution is applied at every lattice site.
In practice, $W$ is usually taken to be non-zero only inside a small window $-K \leq k_1, k_2 \leq K$, where $K$ tends to be rather small indeed (often just one or two).
Since $K$ represents the largest distance over which correlations can be encoded into the trainable weights of any given convolutional layer, theories with long correlation lengths $\xi \gg K$ require convolution-based models to be sufficiently \textit{deep} in order to indirectly model correlations over large scales.

Rather than a number of nodes $H$ in the hidden layers of a fully-connected network, we may specify a number of input and output `channels', each of which has its own convolutional kernel.
Since we are dealing with a single-component scalar field, the first layer will have one input channel.
The number of output channels will depend on how many numbers are required to parametrise the transformation of each degree of freedom, i.e. one for additive layers, two for affine layers, and $3K-1$ for the splines.
Note that, unlike the fully-connected case, the geometry of the objects being convolved must remain intact; one cannot simply pass the passive elements $v^P$ into a convolutional network as an arbitrarily-ordered one-dimensional vector, as we have previously been doing.
Following Reference~\cite{Albergo2021}, we pass two-dimensional configurations into the convolutional networks, but with zeroes as placeholder values for the active partition to maintain the diagonal Jacobian structure of the coupling layer.

Let $M_{\Lambda^P}$ denote an $L \times L$ matrix with ones at the positions corresponding to the \textit{passive} partition, and zeroes at the \textit{active} lattice sites.
Further, let $c_i = 1, \ldots, H_i$ label the channels in the $i$-th layer of the network, and $\sigma$ and $b^{(i)}$ once again represent activation functions and vectors of biases, respectively.
The depth-$D$ convolutional networks tested here can then be written as

\begin{widetext}
\begin{equation} \label{eq:convnet}
    \mathbf{N}_{c_D}(v) = \sum_{c_{D-1}=1}^{H_{D-1}} W_{c_D, c_{D-1}}^{(D)} \star \sigma \Bigg( \ldots
    \sum_{c_1=1}^{H_1} W_{c_2, c_1}^{(2)} \star 
    \sigma \Big( W_{c_1}^{(1)} \star ( M_{\Lambda^P} \odot v ) + b^{(1)}_{c_1} \Big)
    + b^{(2)}_{c_2} \ldots \Bigg) + b^{(D)}_{c_D} \, ,
\end{equation}
\end{widetext}

In the present work we have limited the scope of our quantitative study to models using fully-connected networks as described by Equation~\eqref{eq:mlp}, supplying only indicative examples of models using convolutional networks as described above.

\subsection{Enforcing equivariance} \label{sec:flows:z2}

Convolutions are frequently favoured over linear transformations because they possess the very desirable property that translations of the inputs induce nothing more than translations of the outputs, a trait often referred to as \textit{equivariance} (with respect to translations).\footnote{The translational equivariance of the convolutional networks described by Equation~\eqref{eq:convnet} deserves comment. These networks are really equivariant with respect to the discrete group of translations that are isometries of the checkerboard \textit{sub}-lattices. A consequence of the checkerboarding is that we sacrifice equivariance under the full group of lattice translational isometries.}
Hence, convolutional networks are not required to `learn' that inputs related by a global translation should be considered equivalent.

In addition to the symmetries of the lattice, $\phi^4$ theory possesses a $\mathbb{Z}_2$ symmetry corresponding to invariance of the action under a global sign-reversal of the field,

\begin{equation}
    p(-\phi) = p(\phi) \, .
\end{equation}

Given that equivariant maps are those which commute with the symmetry transformation, it is not difficult to show that, for the $\mathbb{Z}_2$ symmetry, the equivariant maps are \textit{odd} functions, and that equivariance of $f_\theta$ requires the coupling layers $g_i(v_i)$ to be equivariant, meaning that the transformations satisfy

\begin{equation} \label{eq:symm:coupling}
    C_{i,x} \big(-v^A_{i,x} \, ; \, \mathbf{N}_{i,x} ( -v^P_i ) \big) = -C_{i,x} \big( v^A_{i,x} \, ; \, \mathbf{N}_{i,x} ( v^P_i ) \big) \, .
\end{equation}

In our models, $\mathbf{N}_i(v^P)$ are (almost always) fully-connected feed-forward networks as defined by Equation \eqref{eq:mlp}, which are odd functions if we drop the biases and use odd activation functions (e.g. $\tanh$) \cite{Nicoli2020}.
If we make these choices for the neural networks in the affine coupling layers, $\mathbf{s}_i$ and $\mathbf{t}_i$, then Equation \eqref{eq:symm:coupling} is trivially satisfied by implementing one additional step, that is to take the absolute value of the output of the $\mathbf{s}_i$ network.

Enforcing $\mathbb{Z}_2$-equivariance in the RQS transformations is less straightforward; the terms in Equation \eqref{eq:rqs_coupling} cannot all be simultaneously odd.
We implemented a rather crude workaround that involves splitting the batch of latent variables according to $\sgn \sum_{x\in\Lambda} z_x$ (i.e. the initial `magnetisation' of each configuration), and treating the two groups slightly differently within the transformation.
The key observation is that if we take $a$ as 0th knot (instead of $-a$) and construct the spline in the reverse direction, then this is equivalent to taking $C^\text{rqs}_{i,x} \mapsto -C^\text{rqs}_{i,x}$.
Practically speaking, it is simpler to simply reverse the ordering of the $k$ indices in the network outputs $\mathbf{h}_i$, $\mathbf{w}_i$ and $\mathbf{d}_i$ for one of the two groups.

Unfortunately, although the equivariance condition is satisfied, this approach is not entirely legitimate, since the result is a transformation that is not a continuous function of the inputs, thereby failing to satisfy the conditions required for Equation~\eqref{eq:ch_var_formula}.
To see this, observe that $C_{i,x}(v_{i,x} \, ; \, \mathbf{N}_{i,x}(v_i^P))$ may take two possible values for a fixed $v_{i,x}$ and $v_i^P$, experiencing a discontinuous jump when changes in the \textit{active} partition cause the overall magnetisation of $z$ to flip sign.
So, while we include results using these `equivariant splines', we do not recommend using this prescription in future.

It is worth bearing in mind that enforcing symmetries is not absolutely necessary; the Metropolis-Hastings algorithm is guaranteed to converge to the correct target, and therefore reproduce all of its symmetries, as long as the transition kernel is ergodic.
We remind the reader that a sufficient condition is $\tilde{p}(\phi) > 0 \; \forall \phi \in \mathbb{R}^{|\Lambda|}$ \cite{Tierney1994} (so that every configuration has a finite probability of being generated), and that this is guaranteed (for a sensible choice of $r(z)$) since $f_\theta$ is a bijection.
Nevertheless, a guiding principle of optimisation is that it is generally more efficient to enforce known constraints by construction, and benefits of doing so for normalizing flows have been reported in References~\cite{Kohler2020,Kanwar2020,Boyda2020}.


\section{Related work} \label{sec:related}

The first demonstration of a normalizing flow forming the basis of a sampling algorithm for lattice field theory was provided by \citeauthor{Albergo2019} \cite{Albergo2019} for two-dimensional $\phi^4$ theory, using the \textit{Real NVP} architecture described in Section \ref{sec:flows:affine}.
Still with $\phi^4$ as the target theory, \citeauthor{Nicoli2020} \cite{Nicoli2020} used an even more bare-bones flow where the coupling layers simply shift the field variables in such a way that the symmetry under $\phi\to-\phi$ is preserved.
Our work draws on ideas from both of these studies, though we pivot in the opposite direction with respect to Reference \cite{Nicoli2020} by using coupling layers that are \textit{more} flexible than those in \textit{Real NVP}.
More recent work along these lines has been undertaken by \citeauthor{Hackett2021}, who compared several optimisation strategies for flow-based sampling from bimodal distributions, including $\phi^4$ in its broken phase.

Further progress has mostly been on the side of developing the necessary machinery to apply these ideas to lattice \textit{gauge} theories; specifically, those that are invariant under \text{local} $\mathrm{U}(N)$ or $\mathrm{SU}(N)$ transformations.
\citeauthor{Rezende2020}~\cite{Rezende2020} explored several possible approaches to using normalizing flows in cases where the field variables are defined on an $n$-sphere or $n$-torus.
A procedure for constructing normalizing flows that are equivariant under gauge transformations was initially developed by \citeauthor{Kanwar2020}~\cite{Kanwar2020} for the $\mathrm{U}(1)$ case and then extended to $\mathrm{SU}(2)$ and $\mathrm{SU}(3)$ by \citeauthor{Boyda2020} \cite{Boyda2020}.
By definition, a gauge-equivariant flow is one that commutes with the action of the gauge group, which implies that gauge \textit{in}variance is preserved by the flow.
Hence, representative samples of gauge fields can be generated using latent variables drawn from the uniform (Haar) measure for the gauge group, and passing them through a gauge-equivariant flow.
More recently yet, \citeauthor{Albergo2021b} developed the flow-based approach to sampling from theories with dynamical fermions.
A code-based introduction to these methods has been provided by \citeauthor{Albergo2021} \cite{Albergo2021}, which we made use of when implementing convolution-based flow models.

Another recent and highly relevant contribution was made by \citeauthor{Lawrence2021} \cite{Lawrence2021} who showed that, in certain cases at least, it is possible to use a normalizing flow to sample from a theory possessing a `sign problem', which is to say the action is complex and $\exp(-S)$ cannot be interpreted as a measure of probability.

Several alternative ideas that involve training parametric models to perform collective updates predate the use of normalizing flows.
For example, in the `self-learning Monte Carlo' method \cite{Liu2017a} the parametric model describes an effective action for a spin system with $n$-th nearest neighbour interactions whose couplings have been inferred from pre-generated training data, which can then be used to generate Wolff cluster updates \cite{Wolff1989}.
Restricted Boltzmann machines (RBMs) have been embedded in traditional MCMC algorithms \cite{Tanaka2017,Huang2017}, though training the RBM requires pre-generated configurations and its sampling procedure (Gibbs sampling) introduces its own autocorrelation.
Generative adversarial networks (GANs), which can be more flexible than normalizing flows but for which $\tilde{p}(\phi)$ is defined implicitly and cannot be directly computed, have also been used to generate candidate field configurations, but require a lot of additional machinery on top of the GAN itself to ensure that the distribution being sampled from is close to the correct one \cite{Liu2017b,Urban2018}, or otherwise estimate the discrepancy \cite{Singh2020}.

While there have been substantial advances in the use of machine learning to extract physical information for lattice field theories, the generation of samples from some approximation of the true path integral has generally come as an add-on when the tool being used is a generative model.
In contrast, the key strength of normalizing flows is the explicit and tractable density $\tilde{p}(\phi)$ which makes exact sampling possible using the Metropolis test.


\section{Experimental setup} \label{sec:setup}

\subsection{Field theory and observables} \label{sec:setup:phi_four}

For the main part of our study we used the following action:
\begin{equation} \label{eq:phi_four:action}
    S(\phi) = \sum_{x\in\Lambda} \bigg[
    -\beta \sum_{\mu=1}^2 \phi_{x+e_\mu}\phi_x + \phi_x^2 + \lambda(\phi_x^2 - 1)^2 \bigg] \, ,
\end{equation}
which describes a discretised analogue of two-dimensional scalar $\phi^4$ theory with dimensionless couplings $\beta$ and $\lambda$, defined on a periodic lattice $\Lambda$, using $e_\mu$ to denote a unit lattice vector in the $\mu$-th dimension.
Experiments with the non-interacting theory used the `standard' action described in Appendix \ref{sec:appen:phi_four}, which is given by Equation \eqref{eq:phi_four:standard_lattice_action} with $g_0 = 0$.
We focus on isotropic lattices with $6^2 \leq |\Lambda| \leq 20^2$ sites.

A nice feature of the parameterisation given above is that the limit $\lambda \to \infty$, $\phi^2 \to 1$ is very clearly identified as the Ising model at temperature $T \equiv \beta^{-1}$.
Indeed, in the continuum limit $\phi^4$ theory belongs to the Ising universality class, with spontaneous breaking of the $\phi\mapsto-\phi$ symmetry occurring along a critical line $\big(\lambda, T_c(\lambda)\big)$ in the space of couplings.
Defining the reduced temperature $t = \frac{T - T_c(\lambda)}{T_c(\lambda)}$, the asymptotic behaviour of observables as the system approaches criticality is described by power-law dependence on $t$.
For example, the magnetic susceptibility diverges as $\chi \sim t^{-\gamma}$, and the correlation length diverges with a different critical exponent, $\xi \sim t^{-\nu}$.
Eliminating $t$, we see that $\chi \sim \xi^{\gamma/\nu}$.

In a finite volume observables depend on both the couplings and the system size in a non-trivial manner, and their behaviour in the critical region is described by finite-size scaling.
For example, the susceptibility in a volume of linear extent $L$ can be written in the following manner,
\begin{equation}
    \chi = \xi^{\gamma/\nu} g_\chi(L/\xi) \, ,
\end{equation}
in which finite-volume effects have been bundled into a dimensionless scaling function $g_\chi(L/\xi)$, which we notice must tend towards a constant value as $L\to\infty$ and approach $(L/\xi)^{\gamma/\nu}$ for $L \ll \xi$ so as to act as a cutoff.

We now need to specify how we actually measure observables on the lattice.
The basic building blocks are the two point correlation function,
\begin{equation}  \label{eq:phi_four:correlator}
     G(y) = \frac{1}{|\Lambda|} \sum_{x\in\Lambda} \, 
     \Big\langle \big( \phi_{x+y} - \langle\phi\rangle \big)
     \big( \phi_x - \langle\phi\rangle \big) \Big\rangle \, ,
\end{equation}
and its Fourier transform,
\begin{align}
    \tilde{G}(q) &= \sum_{y\in\Lambda} e^{i q \cdot y} G(y) \, .
\end{align}
We have used the translation invariance of Equation~\eqref{eq:phi_four:action} to take a volume-average in Equation~\eqref{eq:phi_four:correlator} for the simple reason that it improves the statistics.

The susceptibility is identified with $\tilde{G}(0)$, but in the classical spin setting it is often expressed in terms of the magnetisation $M(\phi) = \sum_{x\in\Lambda} \phi_x$,
\begin{equation}
    \tilde{G}(0) \equiv \chi = \frac{1}{|\Lambda|} \Big\langle \big(M - \langle M \rangle\big)^2 \Big\rangle \, ,
\end{equation}

Estimators for these observables are easily obtained by exchanging $\langle . \rangle$ for a sample mean, and uncertainties estimated using the bootstrap method \cite{Efron1979,Efron1986}.
However, without explicitly breaking the $\mathbb{Z}_2$ symmetry one will always measure $\langle \phi \rangle = |\Lambda|^{-1} \langle M \rangle = 0$, so if one is interested in the phase transition one can compute separate sample averages for configurations with positive and negative magnetisation, to properly account for the fact that the field variable distribution is bimodal.

The correlation length requires a little more work to measure.
It is the longest mode in the spectrum of $\sum_{x_1 = 0}^{L-1} G(x_1, x_2)$, the correlation function in time-momentum representation, at momentum $q_1 = 0$.
For sufficiently large separations, $x_2$, this takes the form of a pure exponential (a $\cosh$ due to lattice periodicity),
\begin{equation}
    \sum_{x_1=0}^{L-1} G(x_1, x_2) \equiv \hat{G}(x_2) \propto 
    \cosh \bigg( \frac{x_2 - L/2}{\xi} \bigg) \, ,
\end{equation}
from which the correlation length can be extracted through a fit\footnote{
One might hope to fit a sum of exponentials and hence refrain from discarding short separations, but unfortunately this is an ill-conditioned problem \cite{Kaufmann2003}.}
or by computing
\begin{equation} \label{eq:arcosh}
    \xi^{-1}(x_2) = \arcosh \bigg(
    \frac{\hat{G}(x_2 + 1) + \hat{G}(x_2 - 1)}{2\hat{G}(x_2)} \bigg) \, .
\end{equation}
In general, this can be challenging due to low signal/noise ratio at large separations, but with $L \leq 20$ we also suffer from having very few data points to fit.
To slightly improve the situation, we average over the two dimensions when computing Equation~\eqref{eq:arcosh}.

Another option exploits the fact that the lattice propagator takes the form $\tilde{G}(q) \propto \big( \sum_\mu 4\sin^2(q_\mu/2) + \xi^{-2} \big)^{-1}$ in the low-momentum limit, which lets us write \cite{Caracciolo1998}
\begin{equation} \label{eq:low_momentum_xi}
    \xi^2 =
    \frac{1}{2} \sum_{\mu=1}^2 \frac{1}{4\sin^2(\pi/L)} \bigg( \frac{\tilde{G}(0)}{\Real \tilde{G}(\hat{q}_\mu)} - 1 \bigg) \, .
\end{equation}
Here, $\hat{q}_1 = (2\pi/L, 0)$ and $\hat{q}_2 = (0, 2\pi/L)$ are the smallest possible non-zero momenta, and we have used $\tilde{G}(q) + \tilde{G}(-q) = 2\Real \tilde{G}(q)$.

Our intention will be to tune the couplings so as to obtain systems with correlation length $\xi = L/4$, meaning that as we increase the lattice size we are studying essentially the same theory with an increasingly fine resolution.
The purpose of doing this is so that we only see the effect of the number of degrees of freedom on algorithmic efficiency, as we keep the physical size in units of the correlation length constant.
The choice proportionality constant (four) is a reasonable trade-off between the rate at which criticality is approached as we increase $L$, and the size of finite-volume effects contained within scaling functions.
Fixing $\lambda = 0.5$ and allowing $\beta$ to vary, we obtained three separate predictions for the value of $\beta$ that corresponded to $\xi = L / 4$ on the symmetric side of the phase transition.
These values are provided in Table \ref{tab:parameters}.
\begin{table*}
\centering
\begin{ruledtabular}
\begin{tabular}{c c c c c c c c c}
     $L$ & 6 & 8 & 10 & 12 & 14 & 16 & 18 & 20 \\
     $\lambda$ & 0.5 & 0.5 & 0.5 & 0.5 & 0.5 & 0.5 & 0.5 & 0.5 \\
     $\beta$ & 0.537 & 0.576 & 0.601 & 0.616 & 0.626 & 0.634 & 0.641 & 0.645\\
     $\xi$ (fit) & 1.57(2) & 2.05(5) & 2.53(2) & 3.10(8) & 3.40(4) & 4.03(9) & 4.56(5) & 5.1(2) \\
     $\xi$ \eqref{eq:arcosh} & 1.525(3) & 2.005(2) & 2.529(2) & 3.013(3) & 3.471(5) & 3.940(3) & 4.502(6) & 4.903(9) \\
     $\xi$ \eqref{eq:low_momentum_xi} & 1.501(3) & 1.990(2) & 2.524(3) & 3.010(5) & 3.487(8) & 3.970(4) & 4.555(9) & 4.96(1) \\
\end{tabular}
\end{ruledtabular}
\caption{$\phi^4$ couplings and correlation length measurements for the main part of our study. The inverse temperature $\beta$ was tuned such that $\xi \approx L/4$ for each lattice size.}
\label{tab:parameters}
\end{table*}

\subsection{Model details} \label{sec:setup:details}

When investigating the scaling of training costs (Section~\ref{sec:results:scaling}), we used normalizing flows that are a specific hybrid of affine coupling layers and rational quadratic splines, with the parameters of the transformations generated by fully-connected feed-forward neural networks containing a single hidden layer of size $H = |\Lambda|$.
In Section~\ref{sec:results:efficient} we report on the observations that led us to converge on this particular design.

The metric we use to measure the quality of trained models is the average rate at which configurations generated by the model are accepted when used as proposals for a Metropolis-Hastings simulation.
In Section~\ref{sec:results:principle} we verify that this acceptance rate entirely governs the integrated autocorrelation times of the resulting Markov chains, as claimed in Section~\ref{sec:sampling:generative} and specifically Equation~\eqref{eq:tau_rej}.

We used the ADAM optimisation algorithm \cite{Kingma2014} to update the parameters of our models.
The step size or `learning rate' was annealed during training according to a cosine schedule,
\begin{equation} \label{eq:cosine_annealing}
    \eta_t = \frac{\eta_0}{2} \bigg[ 1 + \cos \bigg( \frac{t}{T} \pi \bigg) \bigg] \, ,
\end{equation}
where $T$ is the total number of training iterations.
Note that this learning schedule requires that we specify $T$ before training begins.
There are no additional `stopping criteria'.
The ADAMW variant \cite{Loshchilov2017} along with `warm restarts' \cite{Loshchilov2016} (which amount to resetting $t = 0$) is a useful generalisation which permits us to continue training (perhaps with a larger batch size) if we are not happy with the outcome after $T$ iterations.
After some experimentation with faster initial learning rates, which typically resulted in lower acceptances if the number of training iterations was large, we generally opted for $\eta_0 = 0.001$.

The batch size, i.e. the number of configurations used to estimate the objective function at each training iteration, varied from 250 to 32000 configurations.
In the vast majority of cases the difference between the batch size and the number of training iterations was a factor of one, two or four.
Note that these batch sizes are much larger than those conventionally used in stochastic optimisation.
In fact, it is quite typical to intentionally aim for a highly stochastic trajectory through the space of parameters, by using a very small number of training inputs (as low as 2 in Reference~\cite{Masters2018}) for each update of the model's parameters.
This may seem surprising, particularly since the graphical processing units (GPUs) on which these models are run are entirely optimised for highly parallel computations, so a small batch size is an under-utilisation of these capabilities.
The motivations behind this choice are that the stochasticity reduces the tendency of the model to over-fit the training inputs or otherwise get stuck in local optima \cite{Ge2015,Masters2018}, and tends to find `better' global optima \cite{Zhang2017}.
However, we have no reason to prefer small batch sizes a priori; as explained in Section \ref{sec:flows:training}, the problem of over-fitting training inputs does not apply to us, and we expect the issue of local optima to be alleviated, to some extent, thanks to stochasticity inherited from the random number generator that produces our training inputs.

Unless stated otherwise, one can assume the following for all models presented in the remainder of this paper:
\begin{itemize}
    \item The $\phi^4$ couplings are given by Table \ref{tab:parameters}.
    \item The flow comprises a number of affine coupling blocks followed by a single rational quadratic spline coupling block.
    \item $\mathbb{Z}_2$ equivariance is enforced in the affine and additive coupling layers, as described in Section \ref{sec:flows:z2}.
    \item The splines have 8 segments and do not have $\mathbb{Z}_2$ equivariance enforced.
    \item Neural networks are of the fully-connected kind with have a single hidden layer containing exactly $|\Lambda|$ (i.e. $L^2$) elements, as defined in Equation \eqref{eq:mlp} with $H = |\Lambda|$.
    \item We do not apply an activation function to the output layer of the $\mathbf{s}$ and $\mathbf{t}$ networks in the affine (or additive) layers.
    \item Metropolis-Hastings simulations ran for $10^5$ steps.
    \item In figures, data points and error bars are an average and range taken over three identical models with different random initialisations.
\end{itemize}

Our code, ANVIL~\cite{ANVIL}, is publicly available.
It uses the PyTorch library~\cite{PyTorch} for constructing and training models, and Reportengine~\cite{Reportengine}, a declarative framework for performing scientific analysis.

\subsection{Summary of the procedure}

A training iteration consists of the following steps:
\begin{enumerate}
    \item Sample from Equation~\eqref{eq:prior} to generate a batch of $N$ `latent configurations' --- $\{z^{(1)}, z^{(2)}, \ldots, z^{(N)}\}$ where $z^{(n)} \sim r(z)$ --- with each configuration $z^{(n)}$ comprising $|\Lambda|$ uncorrelated Gaussian numbers.
    \item Pass these variables through the layers of the model, calculating the logarithm of the Jacobian determinant, $\log \left| \partial g_i / \partial v_i \right|$, for each layer $g_i$ as it transforms one of the two partitions. This results in $N$ candidate field configurations and $N$ Jacobian determinants $\log \left| \partial f_\theta / \partial z \right|$ corresponding to the full transformation $\phi = f_\theta(z)$.
    \item Compute the action, Equation~\eqref{eq:phi_four:action}, for the batch of candidate field configurations.
    \item Average the action and Jacobian over the batch, to provide an estimate of the reverse Kullbach-Leibler divergence, Equation~\eqref{eq:kl_estimator_reverse}.\footnote{Strictly speaking we require the \textit{gradient} of Equation~\eqref{eq:kl_estimator_reverse} with respect to the parameters of the model. This is performed automatically by PyTorch's `autograd' machinery~\cite{PyTorch}. Gradients are propagated through neural networks using the backpropagation algorithm~\cite{Rumelhart1986}.}
    \item Update the parameters of the model by a small increment in the direction of steepest gradient using the ADAM or ADAMW optimisation algorithms.
\end{enumerate}

Once we have a trained model, we move onto the sampling.
We generate a large sample of candidate configurations from the model, along with their Jacobian terms, and immediately calculate the quantity $\log w(\phi) = -\log \tilde{p}(\phi) - S(\phi)$ for each candidate configuration.
We are now fully equipped to run a Metropolis-Hastings simulation as described in Section~\ref{sec:sampling}; for the Metropolis test we simply exponentiate $\log w(\phi') - \log w(\phi)$ to obtain the acceptance probability $A(\phi\to\phi')$ (see Equation \eqref{eq:mh_accept} with $q(\phi \mid \phi') = \tilde{p}(\phi)$ and $p(\phi) \propto \exp(-S(\phi))$).


\section{Results} \label{sec:results}

\subsection{Proof of principle} \label{sec:results:principle}

\begin{figure}
        \includegraphics[width=.48\textwidth]{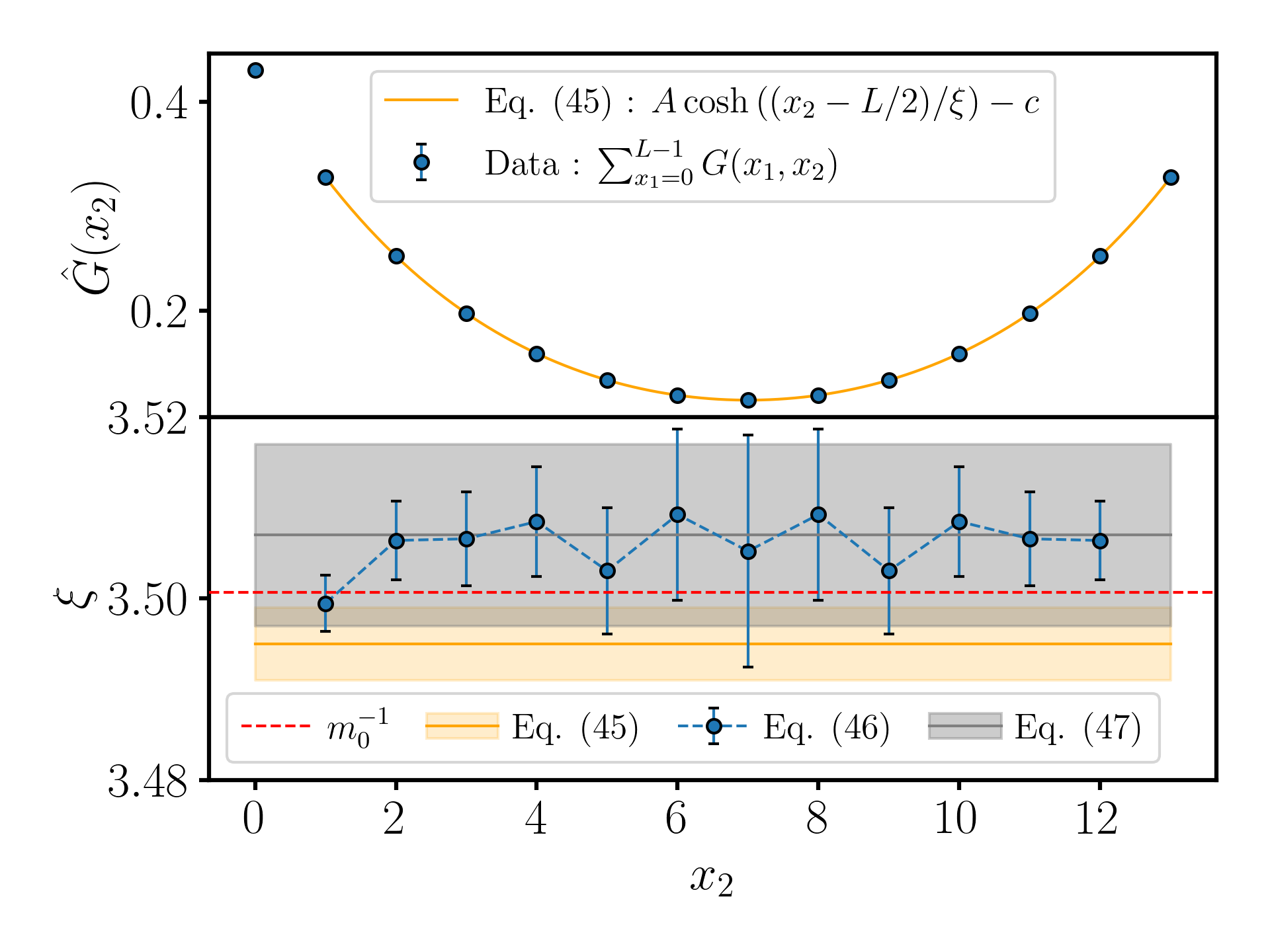}
    \caption{
    Estimates of the correlation length extracted from a sample of $10^6$ field configurations generated by a model trained against Equation~\eqref{eq:phi_four:standard_lattice_action}, the action from a free theory with $L=14$ and $m_0 L = 4$. When this model was used as a generator of proposals for a Metropolis-Hastings simulation, the acceptance rate was 93\%. However, the sample used here is the `raw' output from the model, \textit{without applying the Metropolis test}. Error bars are estimated using the bootstrap method.
    }
    \label{fig:free_theo_corr_length}
\end{figure}

As a basic check that the types of models described in Section \ref{sec:flows} have the capacity to encode the information necessary to trivialize field theories, we trained a set of models to generate free fields.
For this, we found that a sequence of 2--4 blocks of additive transformations performed on equal par with the more flexible affine and spline layers.
Figure \ref{fig:free_theo_corr_length} demonstrates that the candidate field configurations generated by models with very high acceptance rates are indeed representative of the desired field theory.

In this case we know exactly what is required of $f_\theta$; it must perform a rescaling of the latent degrees of freedom followed by a Fourier transform to real space.
However, it would be wrong to suppose that this is a trivial exercise, because the map is built out of a peculiar set of transformations that individually transform half of the degrees of freedom, conditioned on the other half, and the actual transformation learnt by the model does not, and \textit{cannot}, decompose into the simple steps described above.\footnote{We are interested in gaining a deeper understanding of the transformation learnt by the model, but leave this matter to further investigation.}

\begin{figure}
        \includegraphics[width=.48\textwidth]{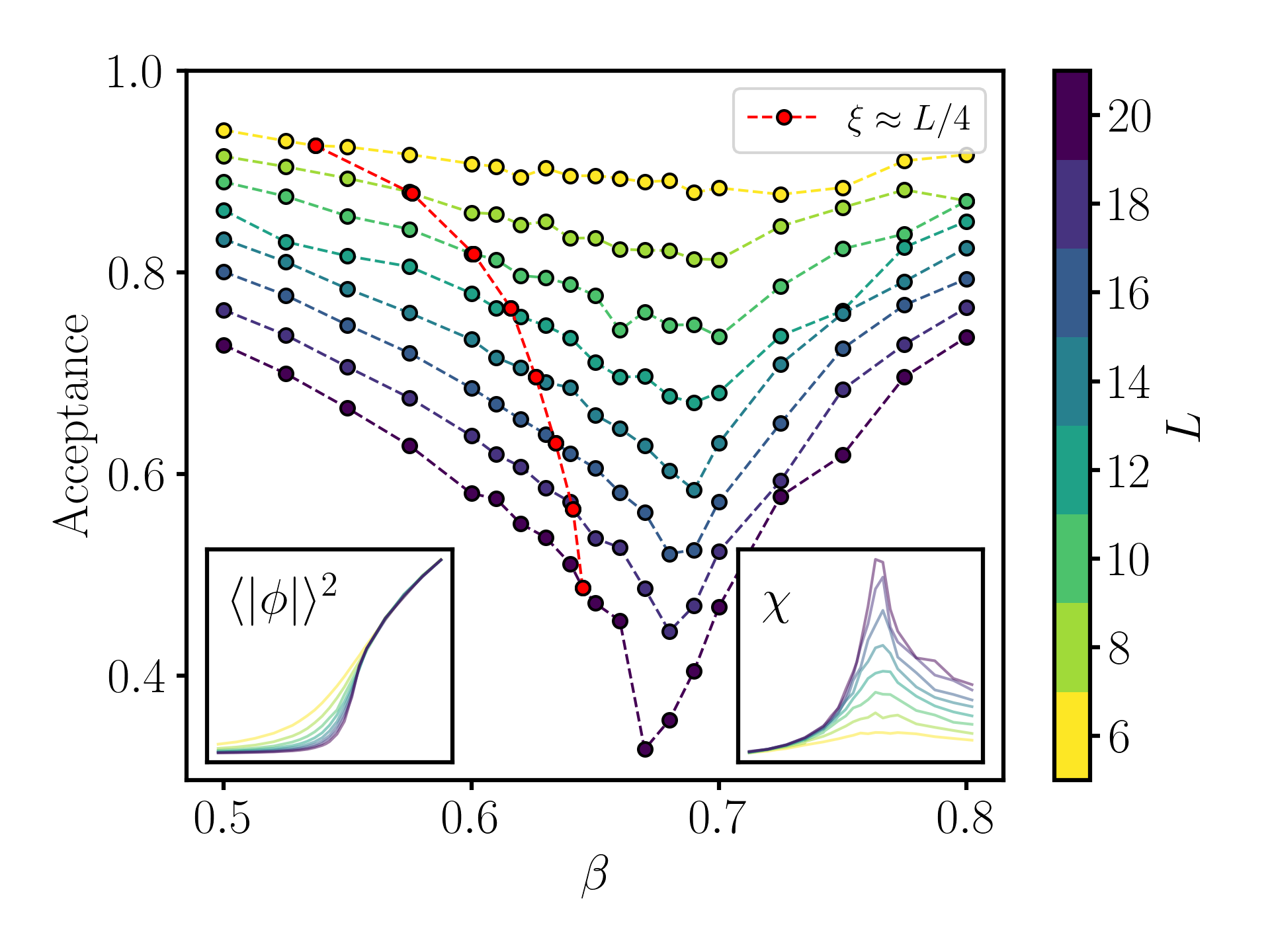}
    \caption{
    Metropolis-Hastings acceptance rates for a set of models trained at different values of the inverse temperature, $\beta$, crossing the critical temperature $(\beta \approx 0.67$ for largest lattice).
    The red points correspond to temperatures used in our main study (values provided in Table~\ref{tab:parameters}. The inset figures show the magnetisation and susceptibility over the same range of $\beta$. Models consisted of a block of affine layers followed by a spline block. The affine layers had $\mathbb{Z}_2$ equivariance enforced as described in Section~\ref{sec:flows:z2}, but this was only true of the spline layers in the low-temperature phase (see Figure \ref{fig:spline_equivar} for explanation). 
    For each lattice size, models were identical (as outlined in Section~\ref{sec:setup:details}) and were trained in an identical fashion (16000 iterations with a batch size of 16000).
    }
    \label{fig:beta_vs_acceptance}
\end{figure}

Moving onto the $\phi^4$ theory, we fixed $\lambda = 0.5$ and trained hybrid affine-spline models at various values of the inverse temperature $\beta$ so as to cross the phase transition.
In this study, emphasis was placed on like-for-like comparison of models trained against different targets, rather than maximising the acceptance rate.
Figure~\ref{fig:beta_vs_acceptance} shows that high acceptance rates are possible in both the symmetric and the broken phase of $\phi^4$ when using a unimodal Gaussian prior.
We also see that, as should be expected, trivializing the theory becomes increasingly challenging as the phase transition is approached.
Though this is still interesting, one should be cautious when interpreting Figure~\ref{fig:beta_vs_acceptance}.
In reality the problem is probably much easier for short correlation lengths than implied by these results, in part because the fully-connected neural networks will contain a high level of redundancy since many degrees are effectively decoupled.

As an aside, we found that this sort of `parameter scan' can be performed efficiently using a single model that is initially trained at a high temperature, by adjusting the temperature over a sequence of training phases; in other words, a model trained at temperature $\beta_0$ can be re-trained at temperature $\beta_0 + \delta\beta$ with relatively little effort.
A thorough investigation into the potential of this feature was recently provided in Reference~\cite{Hackett2021} (see `adiabatic retraining').

\subsection{Acceptance rates and autocorrelation times} \label{sec:results:autocorr}

\begin{figure}
        \includegraphics[width=.48\textwidth]{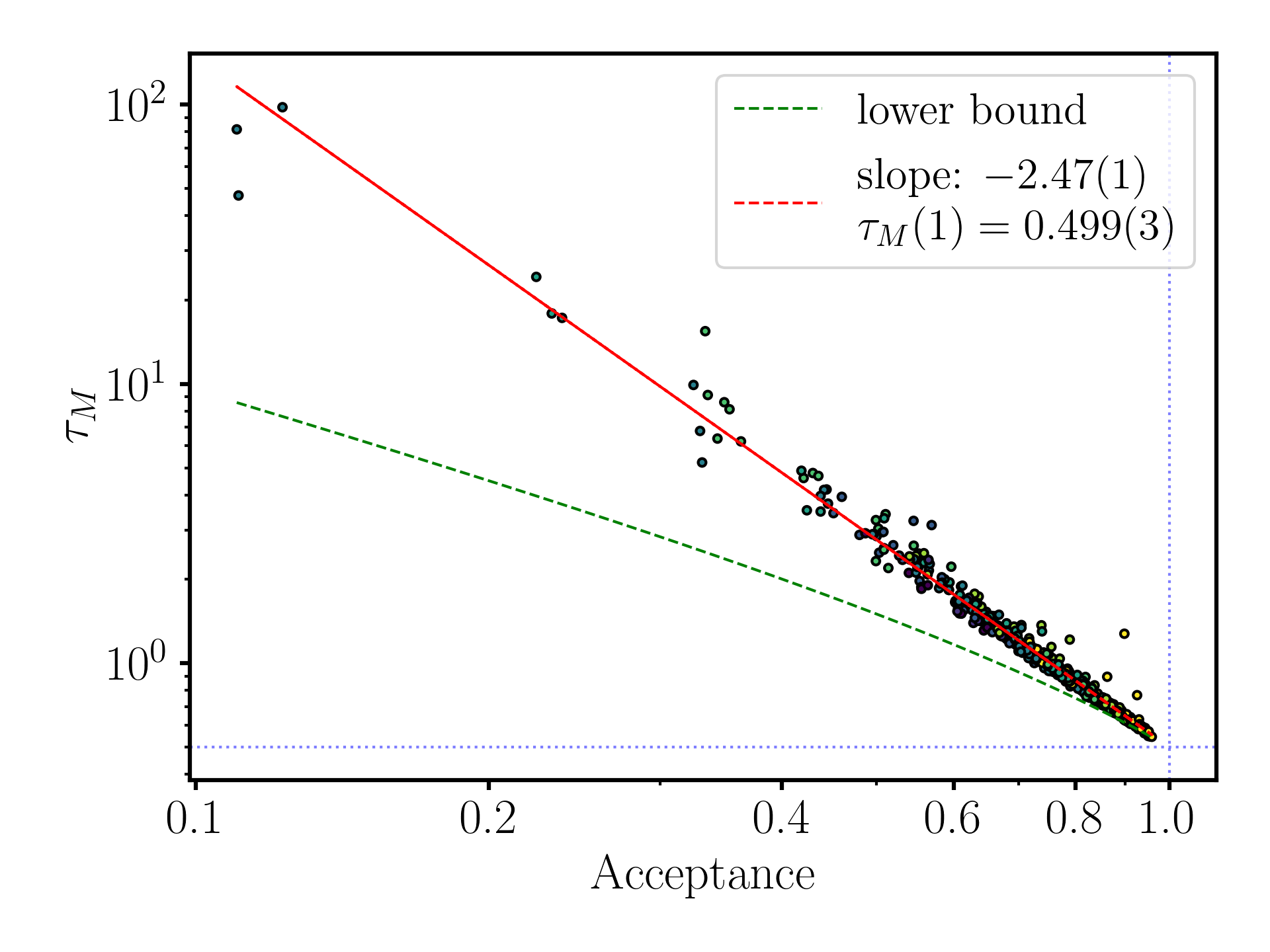}
    \caption{
        Relationship between Metropolis-Hastings acceptance fraction and integrated autocorrelation time.
        Both the gradient and the intercept were left unconstrained in the least-squares fit.
        The theoretical lower bound from Equation \eqref{eq:lower_bound} is also plotted in green, for comparison.
        The data was obtained by sampling from models trained on a broad range of systems, including those specified in Table~\ref{tab:parameters}, those depicted in Figure~\ref{fig:beta_vs_acceptance}, and others trained during various experiments.
    }
    \label{fig:acceptance_vs_tau}
\end{figure}

\begin{table}
\centering
\begin{ruledtabular}
\begin{tabular}{c c c c c c c}
$2 \tau_\obs$ & 1 & 2 & 5 & 10 & 100 & 1000\\
$\E_{\phi \sim p} \E_{\phi' \sim \tilde{p}} \big[ A(\phi\to\phi') \big]$
& 1 & 0.76 & 0.52 & 0.40 & 0.16 & 0.06\\
\end{tabular}
\end{ruledtabular}
\caption{Based on the power-law fit from Figure \ref{fig:acceptance_vs_tau}, $2\tau_\obs$ is the ratio between the effective sample size (that controls the statistical error on observables) and the total length of the Markov chain.}
\label{tab:tau}
\end{table}

\begin{figure}
    \includegraphics[width=.48\textwidth]{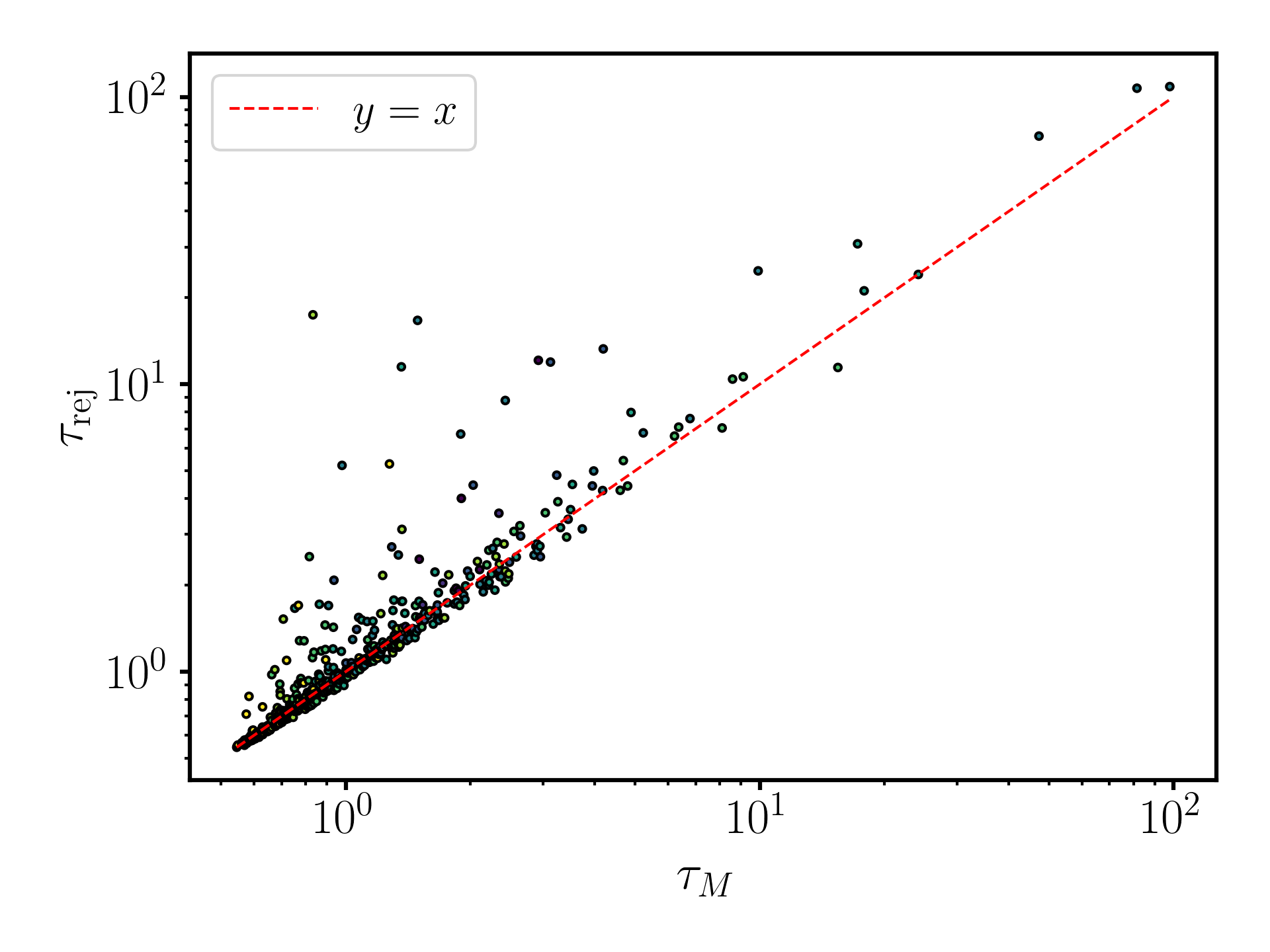}
    \caption{
        Comparison of alternative estimates of the integrated autocorrelation time. On the $x$-axis, $\tau_M$ is calculated in the traditional way, by measuring the autocorrelation function of the magnetisation (see Appendix \ref{sec:appen:autocorr}). On the $y$-axis, the estimator is based on the Metropolis-Hastings rejection rate, using Equations~\eqref{eq:tau_int} and \eqref{eq:tau_rej}.
    }
    \label{fig:tau}
\end{figure}

In Figure~\ref{fig:acceptance_vs_tau} we used the traditional approach to estimating integrated autocorrelation time, based on autocorrelations in the magnetisation of each configuration in the Markov chain, which is described in Appendix \ref{sec:appen:autocorr}.
We now compare this with the observable-independent estimator, $\tau_\text{rej}$, defined by Equations \eqref{eq:tau_rej} and \eqref{eq:tau_int}.
The results, shown in Figure \ref{fig:tau}, show generally good agreement between the two, though where significant discrepancies exist they are always such that the rejection-based estimator returns a larger integrated autocorrelation time than that calculated using the magnetisation.

\begin{figure}
    \includegraphics[width=.48\textwidth]{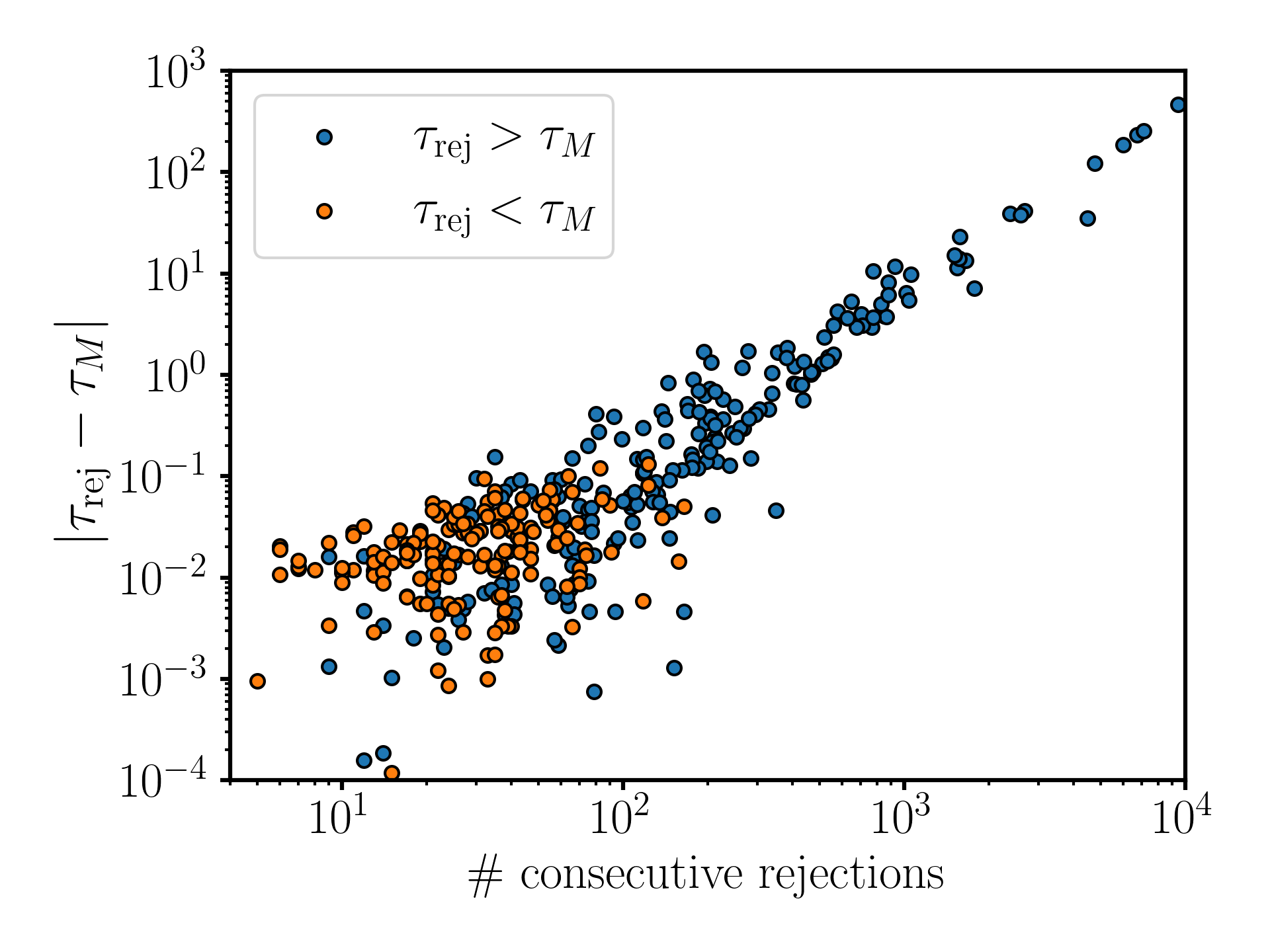}
    \caption{
    Discrepancy between the alternative estimates of the integrated autocorrelation time shown in Figure \ref{fig:tau}, with $x$-coordinates corresponding to the length of the longest consecutive run of rejections occurring in the sampling phase.
    }
    \label{fig:tau_discrep}
\end{figure}

\begin{figure}
    \includegraphics[width=.48\textwidth]{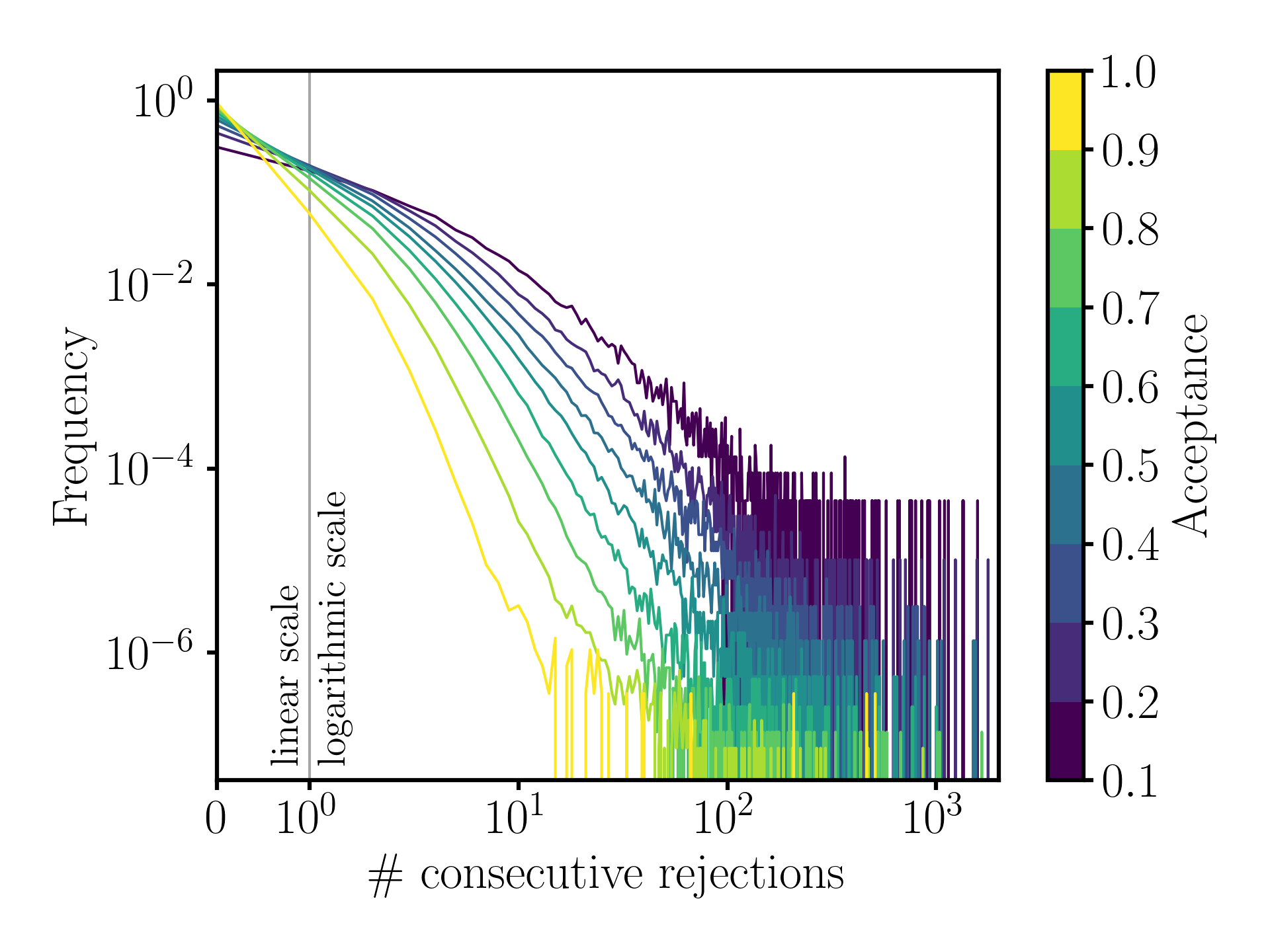}
    \caption{
    Empirical distributions describing the length of periods of consecutive rejections in Metropolis-Hastings simulations involving models with different average acceptance rates. Each empirical distribution combines accept/reject statistics from models with various combinations of layers, trained against different target theories (in terms of lattice size and couplings). The scale on the $x$-axis is logarithmic other than between zero and one (with zero corresponding to instances of two consecutive successful updates). 
    }
    \label{fig:rejection_distribution}
\end{figure}

\begin{figure*}
    \includegraphics[width=.8\textwidth]{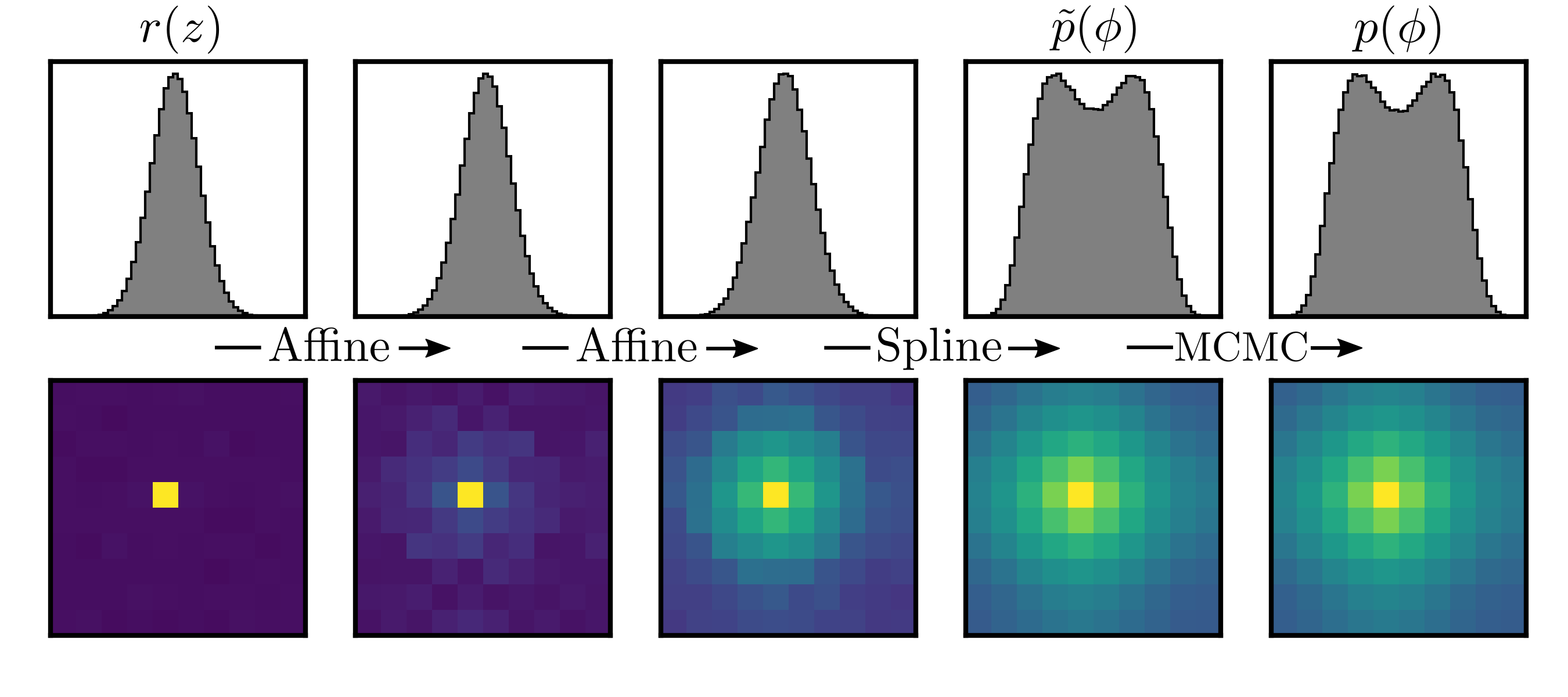}
    \caption{
        Samples of field configurations drawn from each step in the process of generating a representative sample of $\phi^4$ configurations with $\{L, \beta, \lambda\} = \{10, 0.601, 0.5\}$, using a flow model consisting of two affine coupling blocks, followed by a spline.
        The process starts with uncorrelated latent variables, $z \equiv v_1$, and we then sample from the model after each coupling block (layers $i=3, 5, 7$), and finally take the output of the Metropolis-Hastings phase (keeping one in every $2\tau_\text{rej}$ configurations).
        The top row contains histograms of the field variables and the bottom is the two point correlation function from Equation~\eqref{eq:phi_four:correlator}.
        The normalisation used in the colour mapping uses a combination of linear and logarithmic scaling (a symmetric log scaling with a linear threshold of 0.1) to make it easier to see the gradual emergence of correlations.
    }
    \label{fig:spline_flow}
\end{figure*}

In Section~\ref{sec:sampling:generative} it was claimed that, when the proposal distribution $q(\phi' \mid \phi)$ in the Metropolis step, Equation~\eqref{eq:mh_accept}, is \textit{not} conditioned on the current state of the Markov chain and is instead given by $\tilde{p}(\phi)$, the integrated autocorrelation time is determined entirely by the rate at which proposals are rejected or, more precisely, by Equation \eqref{eq:tau_rej}.
Since this has nothing to do with the specifics of the flow model, the system size or the values of the couplings, we simply combined results from a large number of previously trained models (539, to be precise) to verify this property.

Figure \ref{fig:acceptance_vs_tau} provides the necessary empirical evidence that we have indeed nullified any dependence of $\tau_\obs$ on the correlation length of the system, and have therefore eliminated critical slowing down in the sampling phase.
The geometric lower bound on the integrated autocorrelation time from Equation \eqref{eq:lower_bound} is also plotted, but we find that the relationship between acceptance rate and integrated autocorrelation time is fit rather well by a power law,
\begin{equation}
    \tau_\obs \approx \frac{1}{2}
    \E_{\phi\sim p,\phi' \sim \tilde{p}} \big[ A(\phi\to\phi') \big]^{-2.48(1)} \, .
\end{equation}
Table \ref{tab:tau} contains no new information, but rephrases this power-law relation in terms of the acceptance rate required for the effective sample size to be a particular fraction of the Markov chain length.
However, for reasons discussed shortly, we take this scaling relation with a pinch of salt, and advise against extrapolating to lower acceptances and larger autocorrelation times.

As shown in Figure~\ref{fig:tau_discrep}, the discrepancy between $\tau_M$ and $\tau_\text{rej}$ is directly related to the presence of long phases in the Metropolis-Hastings simulation in which every proposed configuration was rejected.
These `rare events', arising from the tails of the distributions depicted in Figure \ref{fig:rejection_distribution}, are problematic when it comes to estimating the integrated autocorrelation time.
The rejection-based estimator is highly sensitive to them and hence picks up a large statistical error (the situation resembles the slow convergence of estimators when local-updates sampling algorithms are required to traverse large energy barriers).
Conversely, the traditional estimator is relatively insensitive to a small number of uncharacteristically long periods of consecutive rejections, which may result in an underestimate of the true integrated autocorrelation time.

In Figure \ref{fig:rejection_distribution} we see the distribution describing the length of runs of consecutive rejections become less long-tailed as the acceptance rate increases.
Although this trend is somewhat obvious in a qualitative sense, the circumstances by which these long-tailed distributions arise are an interesting facet of the scheme used to train these models.
We expand on this in Section \ref{sec:discussion:training}.

\subsection{Finding efficient representations} \label{sec:results:efficient}

The size of models, i.e. the number of trainable parameters, $|\theta|$, will obviously play an important role in determining the scaling of training costs.
Hence, it is not simply a question of building normalizing flows out of highly expressive transformations, but also one of minimising the number of redundant parameters in the model.
More specifically, our ambition must be to find architectures which are able to learn the most \textit{efficient} representations of trivializing maps, for which $|\theta|$ grows slowly as we increase the number of degrees of freedom in the target density.

As stated previously, we found that the flexibility of affine and rational quadratic spline transformations was not put to good use when the target density corresponded to a free theory, with models built from additive layers performing equally well using fewer trainable parameters in total.
We now turn to our main point of focus, which is finding efficiency representations of trivializing maps for strongly interacting theories.
It is a tremendously useful feature of normalizing flows that we are able to sample from the the intermediate layers, which define probability densities in their own right --- see Figure~\ref{fig:spline_flow}.
This gives us insight into the role played by each individual layer.

\begin{figure}
    \includegraphics[width=.48\textwidth]{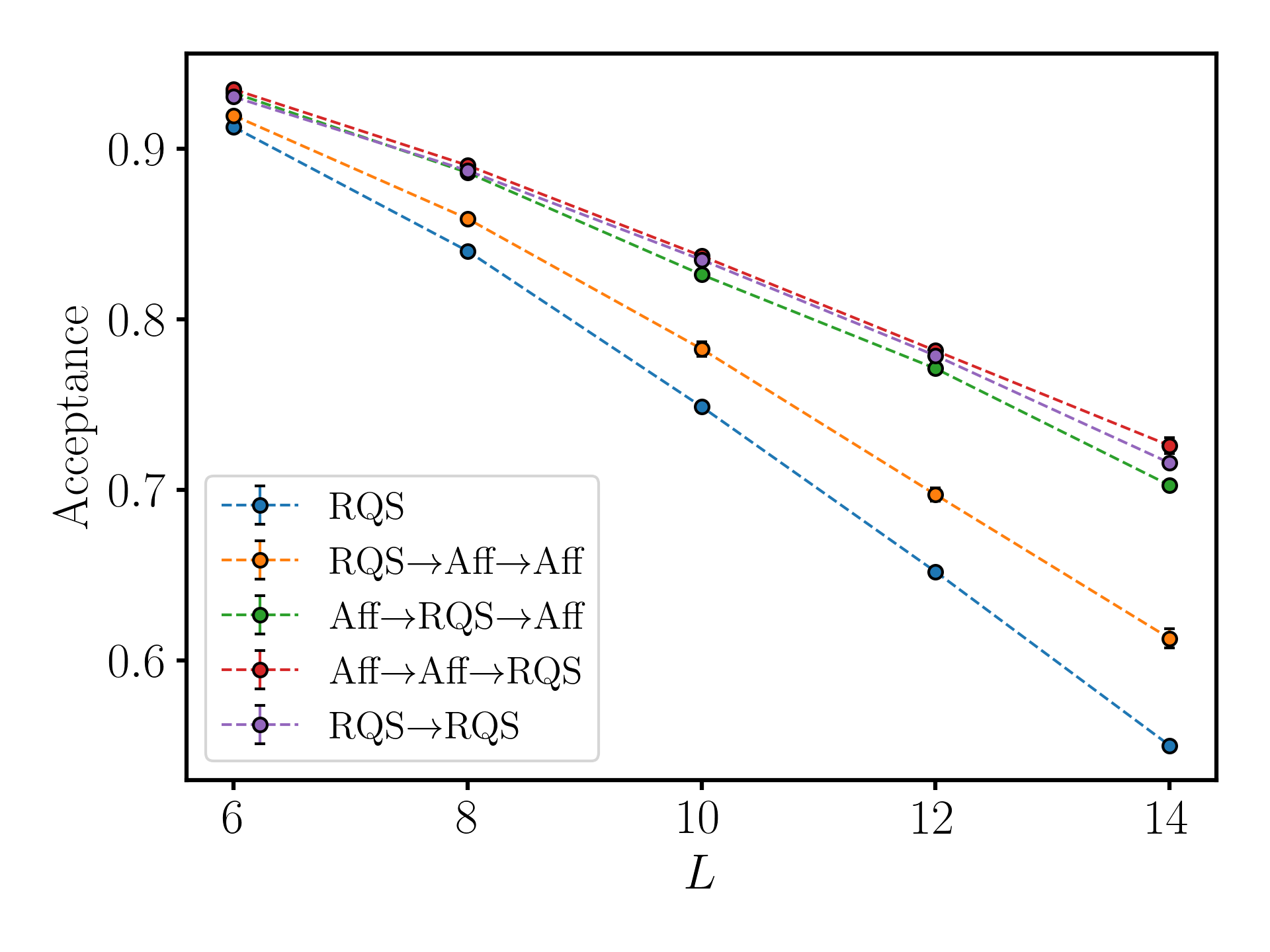}
    \caption{Comparison of flow models with different arrangements of affine (Aff) and rational quadratic spline (RQS) coupling blocks. E.g. the blue model contains a single RQS block, i.e. two RQS coupling layers, whereas the orange model passes the latent variables through an RQS block, then two affine blocks.
    \textit{All} neural networks contained a single hidden layer of size $H = |\Lambda|$.
    All models were trained for 16000 iterations with a batch size of 16000.
    The physical parameters used are those listed in Table~\ref{tab:parameters}.
    }
    \label{fig:layer_ordering}
\end{figure}

The transition from disorder to long-range order in lattice $\phi^4$ theory is characterised by a gradual separation of the initial unimodal probability density into two distinct peaks, corresponding to a positive and a negative net magnetisation.
Therefore, in the regime where the density is bimodal, a normalizing flow transforming Gaussian latent variables must at some point enact a $\mathbb{Z}_2$-symmetric bulk-shifting of probability density to $\pm \langle |\phi| \rangle$.
For flows using solely affine layers, we observed that the task of transforming a unimodal density into a bimodal one was almost always designated to the final coupling block, after earlier layers had resolved the general structure of correlations in the unimodal setting.
We suspect that this is because the functions being modelled by the neural networks simplify when their inputs are distributed unimodally around zero, whereas they must become more strongly non-linear, and hence harder for the neural networks to approximate, when their inputs are distributed bimodally.
When an RQS block was introduced, this always took on the unimodal-bimodal transformation regardless of its position in the flow.
However, Figure~\ref{fig:layer_ordering}, which compares various orderings of affine and spline blocks, shows that the strongest design starts with affine layers and finally applies a single block of spline transformations.

\begin{figure}
    \includegraphics[width=.48\textwidth]{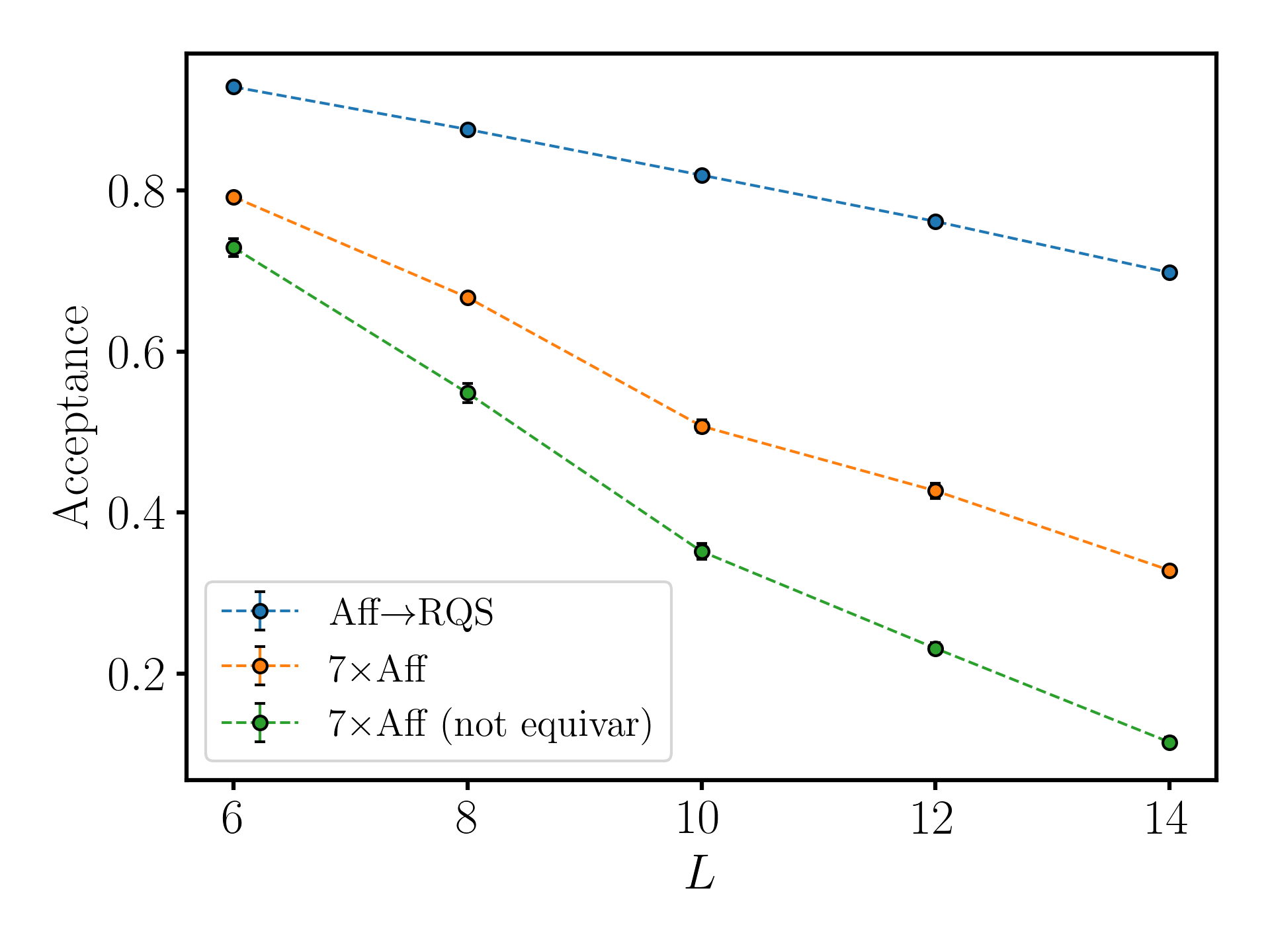}
    \caption{
        Comparison between three groups of flow models using different coupling layers.
        One RQS block contains approximately the same number of parameters as six affine blocks.
        `Equivar' refers to the $\mathbb{Z}_2$-equivariant affine layers described in Section \ref{sec:flows:z2}.
        The green data points correspond to flows in which the affine layers were similar to those used in Reference \cite{Albergo2019} (though ours used neural networks that were both narrower and shallower).
        Models were trained in an identical fashion, for 16000 iterations with a batch size of 16000.
        The physical parameters are given in Table~\ref{tab:parameters}.
    }
    \label{fig:affine_vs_spline}
\end{figure}

Our expectation was that the additional flexibility offered by RQS transformations might help to perform the challenging splitting and shifting of density in a way that is more sensitive to subtle differences in the inputs than is possible with affine layers.
This is confirmed emphatically by the results in Figure~\ref{fig:affine_vs_spline}, for which we took a seven-block affine flow and substituted six blocks of affine coupling layers for a single RQS block in order to compare models with an approximately equal number of trainable parameters.
However, the substitution of many inexpressive affine layers for a single, highly flexible RQS block is a one-off trick; the results in Figure~\ref{fig:layer_ordering} show that adding a second spline block fails to improve the acceptance rate any more than adding another affine block.
It appears that, with the unimodal--bimodal transformation taken care of in the final layer, the remainder of the trivializing map can be modelled more efficiently using the simpler transformations.

\begin{figure}
    \includegraphics[width=.48\textwidth]{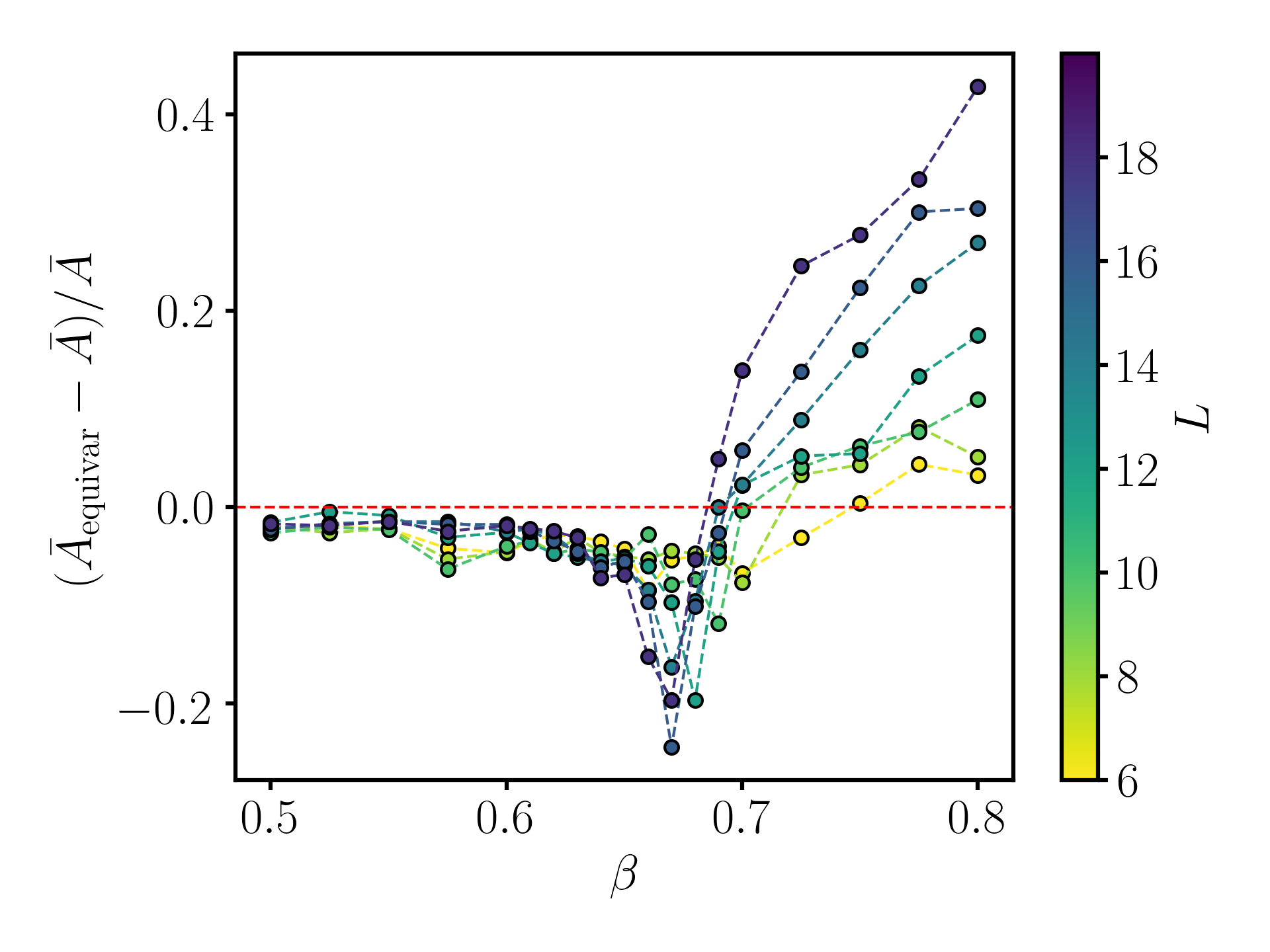}
    \caption{Changes in the Metropolis-Hastings acceptance rate (denoted by $\bar{A}$) of spline-based models, due to enforcing $\mathbb{Z}_2$-equivariance in the spline layers, as described in Section~\ref{sec:flows:z2}. Couplings and training hyper-parameters were identical to those in Figure~\ref{fig:beta_vs_acceptance}. We verified (by inspection of histograms) that each non-equivariant model retained an approximate symmetry, such that both positive and negative magnetisations were sampled with similar frequency.
    }
    \label{fig:spline_equivar}
\end{figure}

From Figure~\ref{fig:affine_vs_spline} we also see that enforcing $\mathbb{Z}_2$-equivariance in the affine layers, as described in Section~\ref{sec:flows:z2}, leads to higher acceptances that reduce less steeply as the lattice size increases.
Despite the theoretical flaw in the design of the $\mathbb{Z}_2$-equivariant spline layers, we still report on a brief investigation into their performance.
Figure~\ref{fig:spline_equivar} shows that in the symmetric phase (which includes the couplings in Table~\ref{tab:parameters}) we did not find that using these layers improved acceptance rates, though they appear to do so in the broken phase.

\begin{figure}
    \includegraphics[width=.5\textwidth]{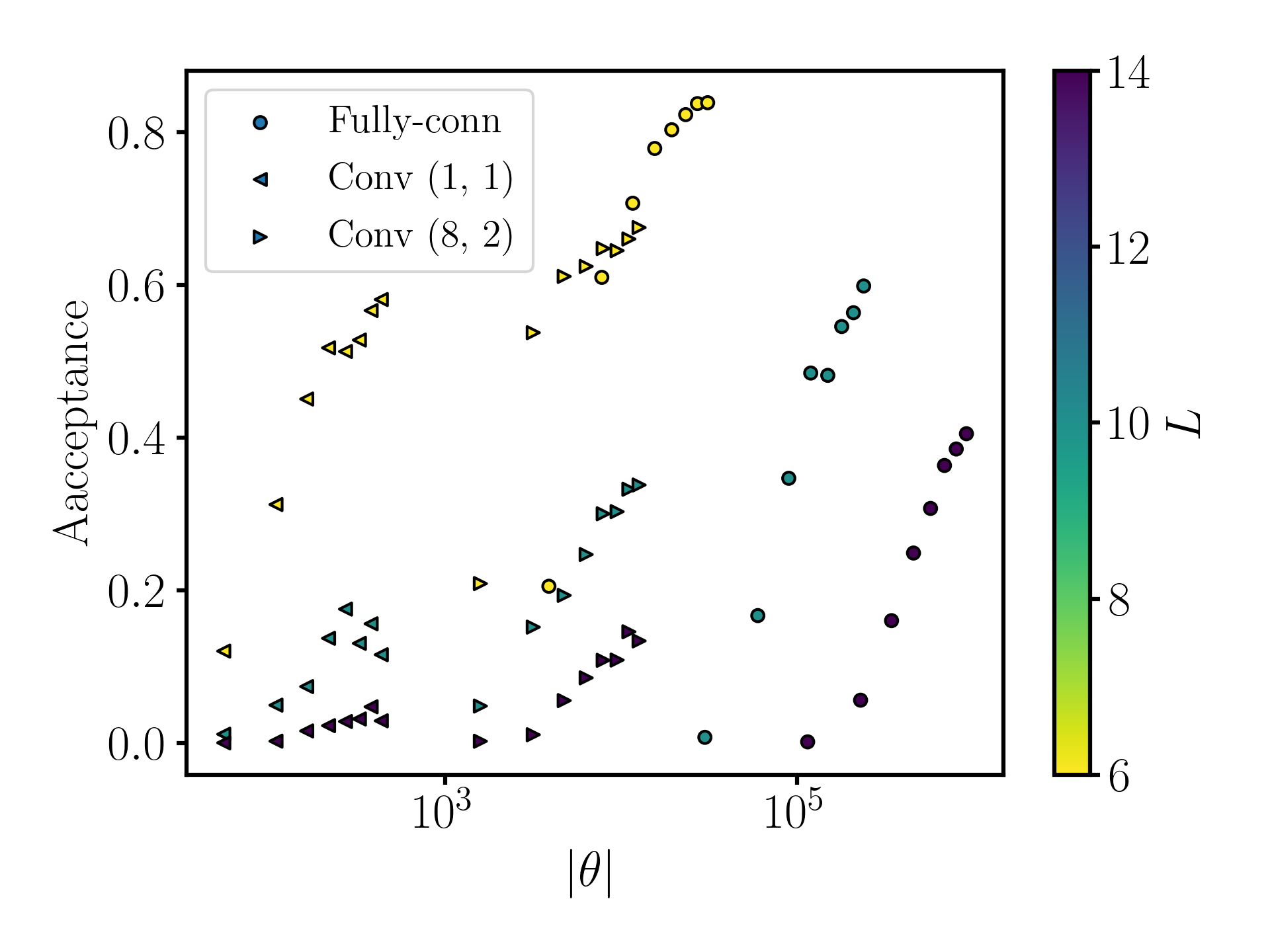}
    \caption{Results from preliminary experiments using \textit{affine} flows with convolutional networks, compared to the fully-connected networks used throughout the rest of this study (single hidden layer of size $\lvert\Lambda\rvert$). The number of parameters, $\lvert\theta\rvert$, various through the number of affine coupling blocks (1--12). Convolutional networks had square kernels of size $3\times 3$. The brackets in the legend denote the number of channels per hidden convolutional layer, and the number of hidden layers, respectively. Models were trained for 16000 iterations with a batch size of 16000. Physical parameters are as in Table~\ref{tab:parameters} for $6^2$, $10^2$ and $14^2$ lattices.
}
    \label{fig:convnet}
\end{figure}

We also report on some preliminary experiments with convolutional networks, summarised in Figure~\ref{fig:convnet}.
The convolution-based models are highly parameter-efficient in comparison with the fully-connected ones, since the number of parameters is decoupled from the lattice size (with the caveat that deeper models are required when the correlation length grows proportionally to the lattice size).
We expect this parameter efficiency to translate to improved scaling of training costs with respect to models using fully-connected networks.
However, on these small lattices we were able to reach significantly higher acceptance rates using fully-connected networks, in less time than it took to train even the simplest convolution-based model.
We are currently working on a more systematic and large-scale comparison of models based on convolutions versus fully-connected networks, focusing particularly on the scaling towards the continuum limit.

Having converged on a general recipe for $f_\theta$ --- some $\mathbb{Z}_2$-equivariant affine layers followed by a single RQS block, using full-connected neural networks --- we have a number of possible ways to make the map more flexible: adding more affine layers; increasing the size of neural networks; increasing the number of segments in the splines.
Our goal is to figure out the extent to which each of these improve the ability of the model to fit $p(\phi)$ while minimising the amount of redundancy in the model's parameters.

\begin{figure}
        \includegraphics[width=.5\textwidth]{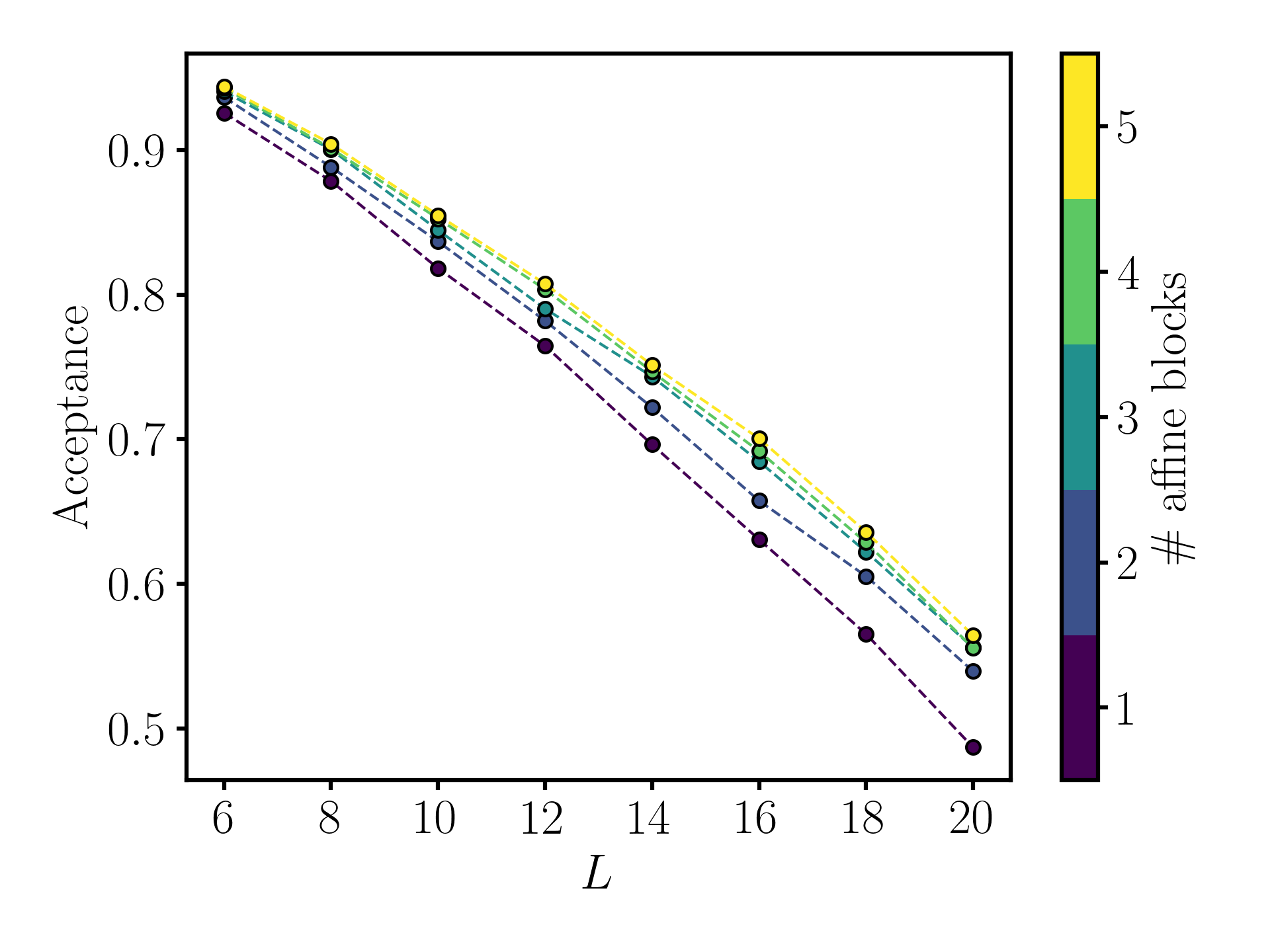}
        \caption{
        Improvements in the Metropolis-Hastings acceptance rate due to adding more affine coupling blocks, before a rational quadratic spline block performs the final transformation. Models were trained for 16000 iterations with a batch size of 16000, and the physical parameters are those found in Table~\ref{tab:parameters}.
    }
    \label{fig:n_affine}
\end{figure}

Firstly, our experiments, shown in Figure \ref{fig:n_affine}, indicate that prepending more affine layers to the flow leads to modest improvements in the acceptance rate for a fixed training length, becoming increasingly worthwhile as the system size increases.
However, there is certainly a law of diminishing returns, and on the lattices that we explored for this work the acceptance seems to plateau by the time we reach five affine layers. 
The law of diminishing returns also applies to adding more segments to a spline transformation; we found that using 8 segments was reasonable, and resulted in RQS layers that were substantially more flexible than affine layers without being too slow to train.

\begin{figure}
        \includegraphics[width=.5\textwidth]{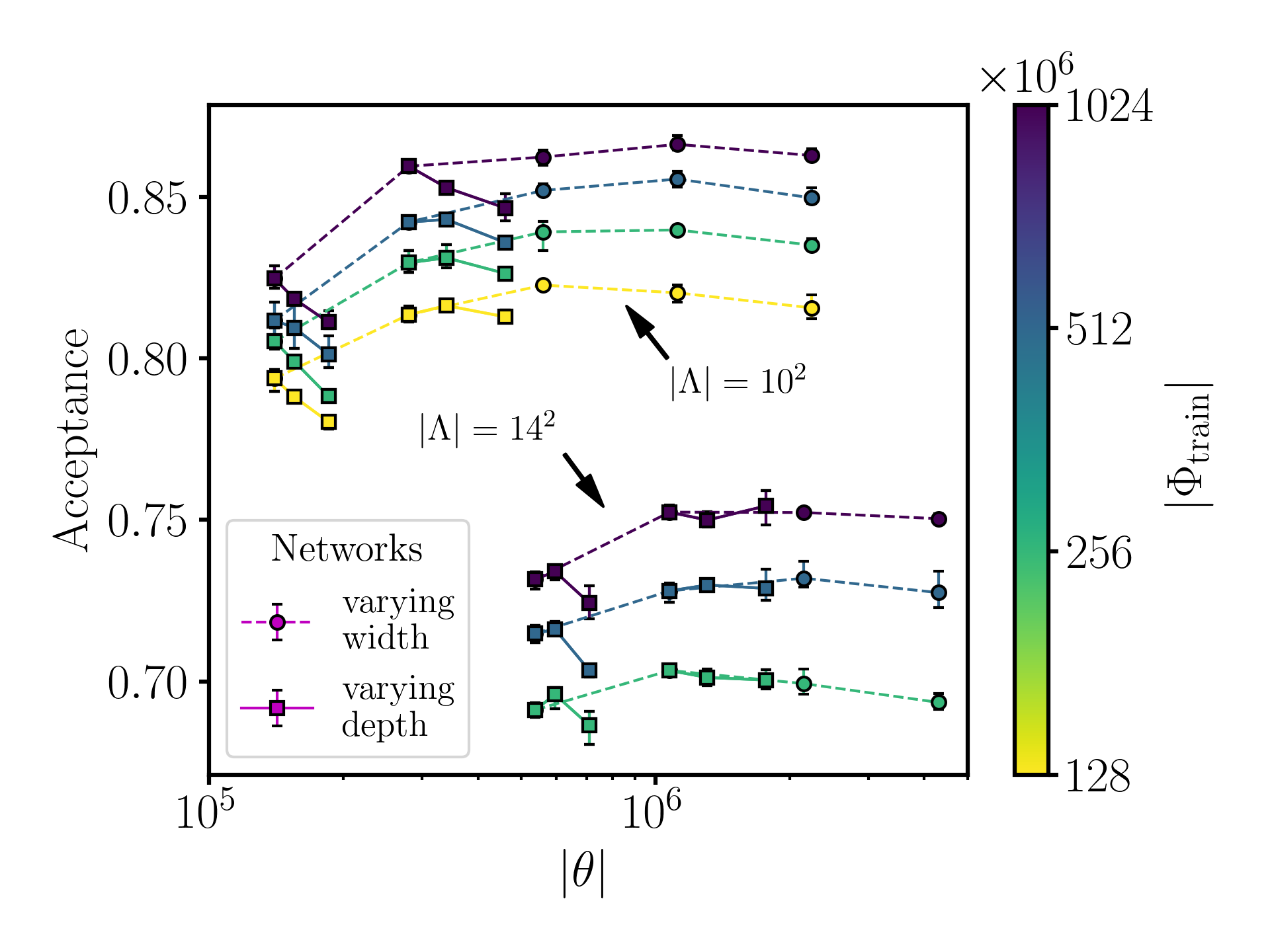}
        \caption{
        Comparison of Metropolis-Hastings acceptance rates for flow models with a different number of trainable parameters, $|\theta|$, within their neural networks.
        The colour axis, $|\Phi_\text{train}|$, denotes the total number of configurations used in the optimisation (the batch size multiplied by the number of training iterations).
        The dashed lines connect models whose neural networks had a single hidden layer, of varying width, whereas following the solid lines equates to increasing the number of hidden layers without changing the widths.
        Models comprised one affine block followed by one RQS block and were trained for 32000 training iterations with a range of batch sizes.
        Physical parameters are provided in Table~\ref{tab:parameters}.
    }
    \label{fig:networks}
\end{figure}

During numerous experiments with deep fully-connected networks (only some of which are reported in Figure~\ref{fig:networks}) we failed to observe any consistent improvements, in terms of the Kullbach-Leibler divergence or the acceptance rate, over an equivalent model using neural networks with a single hidden layer, i.e. exactly as defined in Equation \eqref{eq:mlp}.
We checked that the networks did not suffer from the vanishing gradient problem by training them for at least as many epochs as were required for the gradients from all layers to reach the same order of magnitude.
We also substituted the $\tanh$ activation function for a ReLU which only saturates in one direction, although this meant we were no longer able to enforce $\mathbb{Z}_2$ equivariance within the affine layers, which proved to be a poor exchange.

As shown by Figure \ref{fig:networks}, we also found that the benefits of increasing the width of neural networks quickly diminished once the hidden layer contained more than $H = |\Lambda|$ elements (i.e. twice the size of the input layer, which is the passive partition only).
Since a doubling of the neural network widths in spline layers increases the number of parameters by a factor of approximately $3K|\Lambda|$, incurring a considerable increase in training costs, it was almost never a better investment of resources than increasing the batch size or number of training iterations.

Figure \ref{fig:networks} exemplifies one of the key conclusions of our work --- that the acceptance rate of our models is strongly dependent on the amount of effort we put into training, while being relatively oblivious to the act of adding to the total number of trainable parameters in the model.
In the following section we explore this in a more quantitative fashion.

\subsection{Scaling of training costs} \label{sec:results:scaling}

Neglecting algorithmic or hardware-related factors that contribute to scalability, the cost of training a model up to a given value in some performance metric (e.g. acceptance rate) can be measured in terms of the number of training iterations and the batch size.
We will combine these two factors into a single number, $|\Phi_\text{train}|$, that is the total number of configurations that the model has been exposed to during training.
This happens to also be rather convenient for making comparisons with critical slowing down in traditional algorithms, where the integrated autocorrelation time is proportional to the total number of configurations that must be generated to achieve a target error on expectation values.
See Appendix \ref{sec:appen:albergo} for a set of indicative training times, measured in seconds.

\begin{figure}
    \includegraphics[width=.5\textwidth]{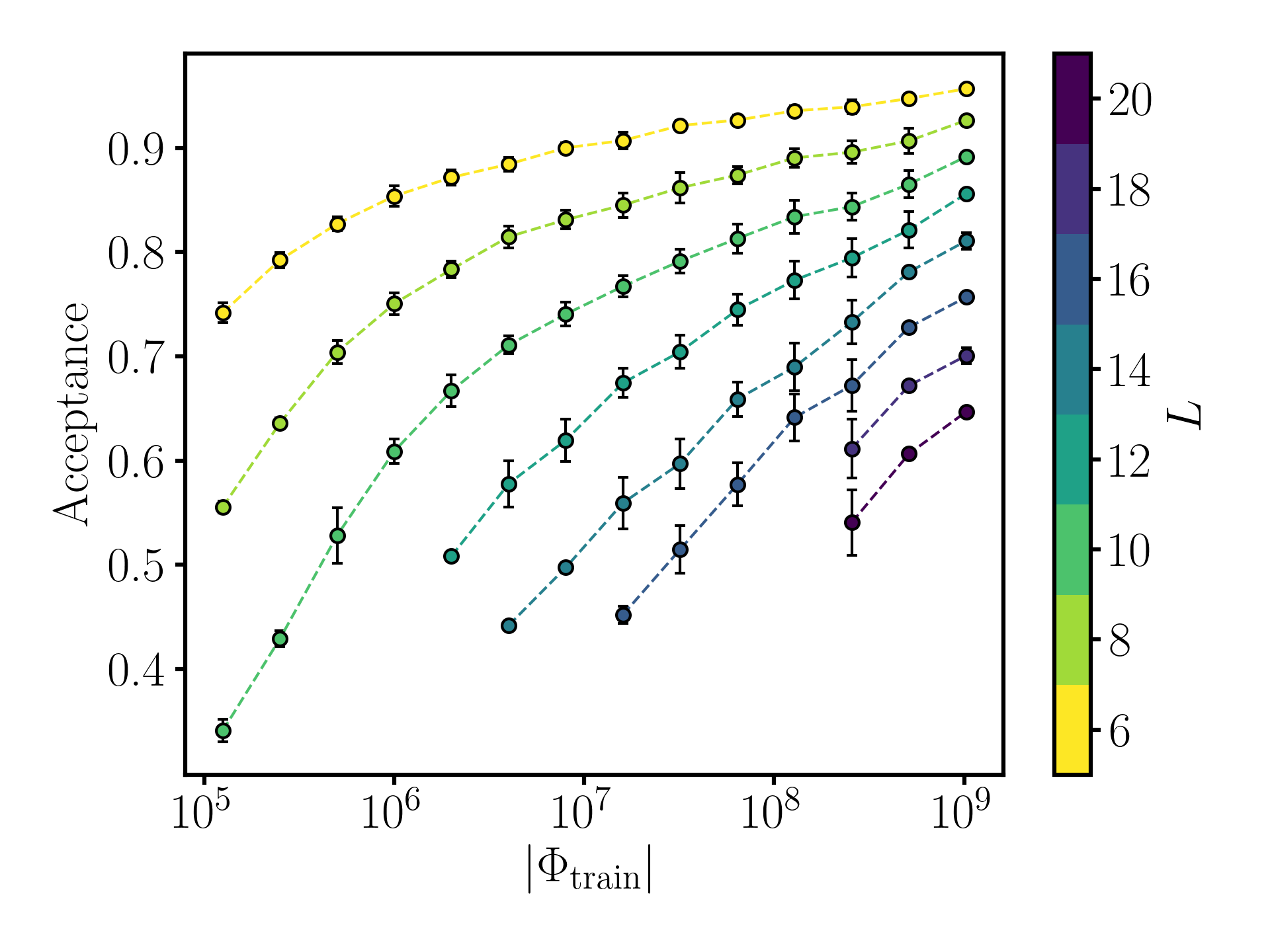}
    \caption{Relationship between the average acceptance rate from fully-trained models and the total number of configurations exposed to the models during training --- i.e. the product of the batch size and the number of training iterations. The $\phi^4$ couplings are given in Table~\ref{tab:parameters}. Error bars are standard deviations over a number of different models trained using different batch sizes ($\leq 32000$) and training lengths ($\leq 32000$), as well as a variable number of affine layers ($\leq 5$).}
    \label{fig:configs_vs_acceptance}
\end{figure}

For this part of the study, we trained a large number of models, as per the recipe of the previous subsection (1--5 affine blocks followed by a single spline), using the $\phi^4$ couplings given in Table~\ref{tab:parameters}.
For training we covered a range of batch sizes (250--32000) and training lengths (500--64000 iterations), both of which were incremented in factors of two.
In Figure \ref{fig:configs_vs_acceptance} we plot the mean and standard deviation of the acceptance, taken over each set of models trained with the same $|\Phi_\text{train}|$.
The error bars are typically much smaller than the difference between adjacent data points, indicating that the overall quality of optimisation is relatively stable under mixing of the batch size and the number of training iterations, provided the total number of configurations from which the model can learn remains fixed.
It is striking that the acceptance rate does not appear to plateau, even for the smallest lattice.
This is a sign that the acceptance is limited not by the expressivity of the model, but by the amount of training (whereas Figure~\ref{fig:affine_vs_spline} shows that pure affine flows are genuinely limited by their expressivity).

\begin{figure}
    \includegraphics[width=.5\textwidth]{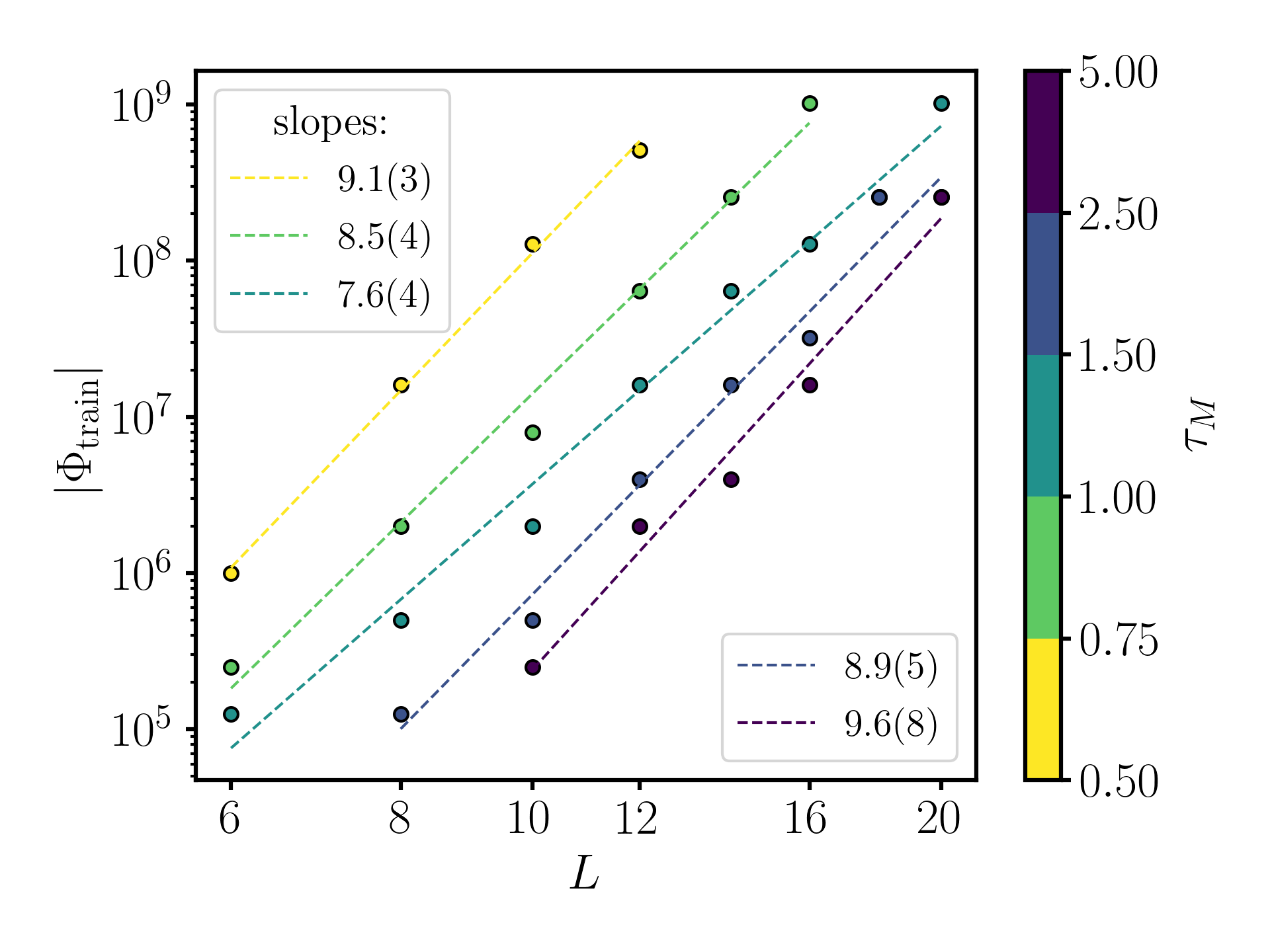}
    \caption{Scaling of the `training cost', measured by the total number of configurations used for optimisation, for models to reach a certain target sampling efficiency.
    Models were sorted into `bins' by the integrated autocorrelation time measured during the sampling phase, and data points represent the best model (lowest $\tau_M$) from each bin.
    As usual, the correlation length was fixed at $\xi \approx L/4$ (see Table~\ref{tab:parameters}).
    Hence, the values in the legend can in some sense be compared to the critical exponent $z_\mathcal{O}$ which determines the critical slowing down in traditional simulations.
    }
    \label{fig:scaling}
\end{figure}

When attempting to quantify the scaling of $|\Phi_\text{train}|$ in a way that can be contrasted with critical slowing down, we found that, in order to get sets of points that could be fit reasonably well using a power law, we needed to group the models by sorting them into bins according to their integrated autocorrelation time, and select the `best' model from each group.
These fits are shown in Figure \ref{fig:scaling}.
The results are sobering; despite the seemingly large improvements compared to the original formulation of Reference \cite{Albergo2019}, the amount of effort required to train these models is growing at an astonishing rate.
While it is true that $|\Phi_\text{train}|$ is a one-off overhead cost, unlike the number of configurations required in a traditional sampling simulation, scaling that goes with the 9th power of the correlation length makes it impossible to avoid the conclusion that this prescription remains far away indeed from a solution to critical slowing down.
In Section~\ref{sec:discussion:csd} we discuss why it is that models with relatively few trainable parameters appear to require such a colossal effort to train.


\section{Discussion} \label{sec:discussion}

\subsection{The most efficient representations are shallow} \label{sec:discussion:shallow}

With few exceptions, machine learning has followed a trend towards increasingly deep architectures, in spite of the fact that many of these architectures, including the fully-connected feed-forward neural networks defined used in this work, require only a single hidden layer to act as universal approximators.
The preference for deep architectures stems from two key principles~\cite{Bengio2009}.
Firstly, deep architectures are less inclined to over-fit training data and hence generalise better than shallow ones.
However, as explained in Section \ref{sec:flows:training}, over-fitting is not a problem we will face since each training iteration exposes the model to a set of previously-unseen training inputs.
Secondly, deep architectures can represent \textit{some} highly nonlinear functions more efficiently than shallow architectures.
This is connected to their ability to represent functions through hierarchical dependencies between abstract `features' (Reference \cite{Lin2016} argues, convincingly, that this is essentially a consequence of the laws of physics).
The truth of this can be rigorously demonstrated for a number of architectures comprising specific functions \cite{Bengio2007,Mhaskar2017,Lin2016}, and empirical evidence is so overwhelming that `deep and cheap' has become somewhat of a mantra in machine learning.

Nevertheless, the question of whether deep architectures are more efficient is task-specific and should be answered through experimentation.
In our case, experiments indicate that increasing the depth of neural networks does not increase the expressive capabilities of the model in a way that translates to a better fit to the target density.
It is possible that this result is not merely a peculiar outcome of building the map out of coupling layers, whose structure is quite unusual; we draw attention to Reference~\cite{Morningstar2017}, in which the conclusion was that a single-layered restricted Boltzmann machine provides a more efficient representation of an Ising system near criticality than any of its deep generalisations.
Furthermore, given a fixed number of trainable parameters, we found that a shallow flow with a more flexible spline layer dramatically outperformed a deeper flow using affine layers, for the interacting theory at least.

\subsection{A closer look at the training algorithm} \label{sec:discussion:training}

Conventional training, via optimising Equation \eqref{eq:kl_estimator} using a fixed training set, penalises the model for under-estimating the density at any point at which there is a training input.
This tends to result in models that are smoothed approximations of $p(\phi)$, spanning the space of training inputs.
If we were to use such a model as the basis of our sampling algorithm, we would expect to see a steady flux of configurations originating from regions of configuration space `between the peaks' in $p(\phi)$, which would typically be rejected by the Metropolis test.
The situation for our models is quite different.
Instead, the characteristic behaviour, which is known as `zero-forcing', is for the optimisation to quickly purge the density from any region of configuration space in which $p(\phi)$ is very small, resulting in models that fit the modes of the target well but frequently underestimate the low-density tails~\cite{Huang2018b}.
When these underestimated regions are eventually sampled from, the probability of transitioning away is suppressed by a factor of $\tilde{p}(\phi) / p(\phi)$ due to the Metropolis test.
Thus, Markov chain histories in which the acceptance rate is typically quite high, but with occasional, `surprisingly' long periods of consecutive rejections, are a generic feature of models trained in this way.

It is enlightening to look more closely at the precise way in which optimisation based on Equation \eqref{eq:kl_estimator_reverse} acts to fit a model to a target density.
For this purpose it will be more convenient to write the Kullbach-Leibler divergence in the following form:
\begin{equation} \label{eq:kl_reverse_discussion}
    \hat{D}_\text{KL}(\tilde{p} \; || \; p) 
    = \E_{\phi \sim \tilde{p}(\phi)} \big[ \log \tilde{p}(\phi) - \log p(\phi) \big] \, ,
\end{equation}
i.e. with the `irrelevant terms' that are neglected in Equation \eqref{eq:kl_estimator_reverse} put back in.
The origin of zero-forcing is clear; if $\log \tilde{p}(\phi)$ is large in some region of configuration space in which $p(\phi)$ is small, then the optimisation will receive a strong gradient signal, because configurations from this region are generated with high frequency and each contributes a large positive term to the objective function.

\begin{figure}
    \includegraphics[width=.48\textwidth]{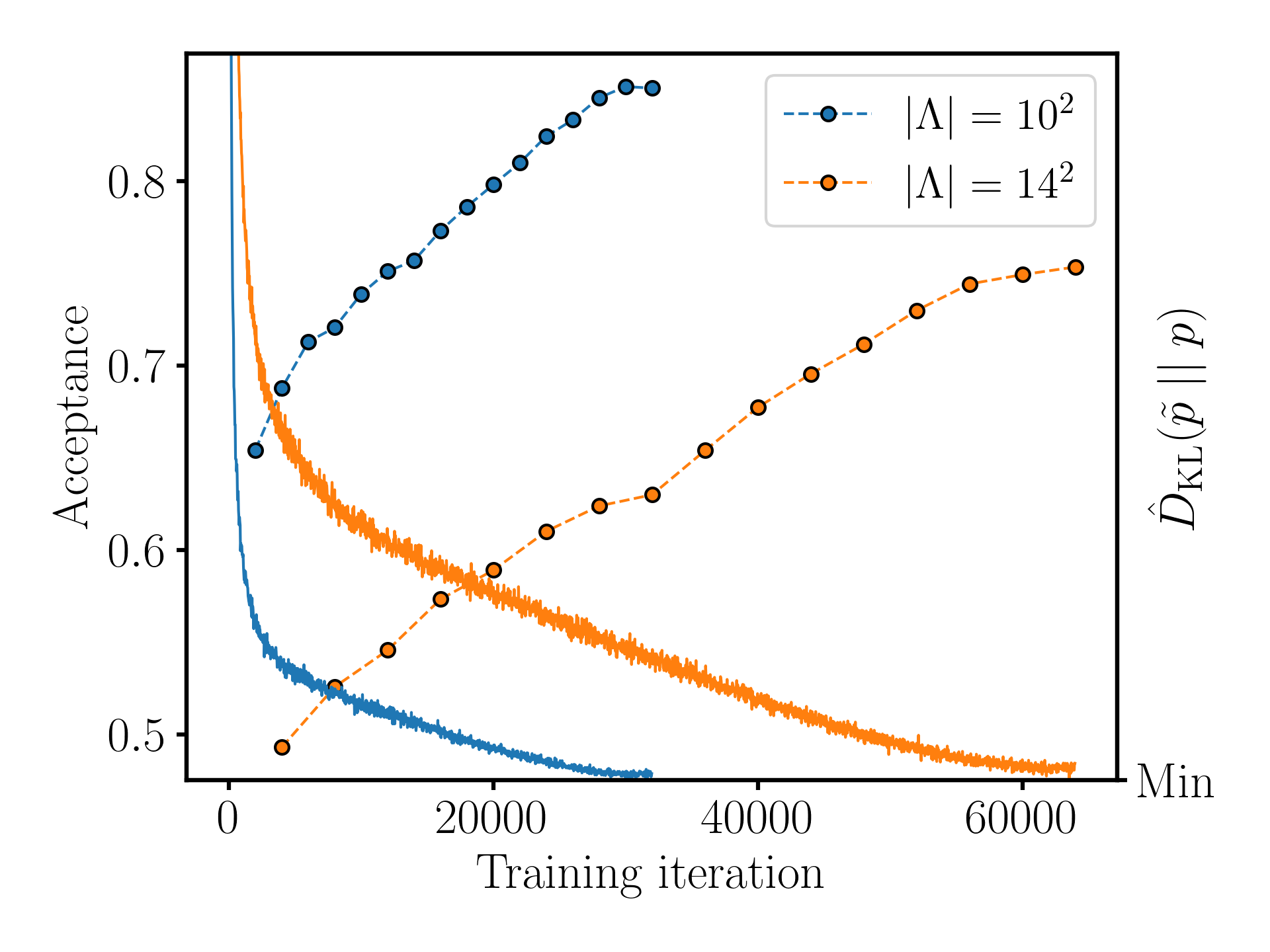}
    \caption{
    Examples of typical training profiles for hybrid affine-spline flow models being trained against interacting theories with correlation length $\xi \approx L/4$. The objective function (right-axis) has been shifted for the purpose of fitting both profiles on one figure.
    The learning rate is being annealed down from $\eta_0$ to zero according to Equation \eqref{eq:cosine_annealing}.
    }
    \label{fig:training_example}
\end{figure}

Using the chain rule, the gradient of Equation \eqref{eq:kl_reverse_discussion} can be written
\begin{equation} \label{eq:gradient_single_config}
    \nabla_\theta \hat{D}_\text{KL}(\tilde{p} \; || \; p)
    = \nabla_\phi \hat{D}_\text{KL}(\tilde{p} \; || \; p) \nabla_\theta f_\theta(z) \, .
\end{equation}
The gradient with respect to generated configurations contains two terms whose roles can be understood intuitively.
One of these acts to reduce the action, $\E_{\phi \sim \tilde{p}(\phi)} \big[ - \log p(\phi) \big]$, driving the optimisation towards the mode(s) of the target density.
This process is `boosted' by the zero-forcing property described previously.
The second term acts to increase the entropy, $\E_{\phi \sim \tilde{p}(\phi)} \big[ - \log \tilde{p}(\phi) \big]$, driving density away from the mode(s), which is necessary to fit the tails of the target.
Figure \ref{fig:training_example} shows a typical training profile, in which the rapid fitting of the modes, boosted by zero-forcing, yields a fairly high acceptance rate early on in the training, after which a more gradual process of expanding around the modes to fit the low-density tails takes over.
Also note that, as the model expands to fill the tails of the target and and the acceptance rate goes up, we also see that the distribution of consecutive rejections becomes less long-tailed, as demonstrated by Figure \ref{fig:rejection_distribution}.

\subsection{A small warning regarding ergodicity} \label{sec:discussion:warning}

It is worth considering the expected outcome of training a model that is seriously deficient in its inherent capacity to approximate the target density.
In such a situation Equation \eqref{eq:kl_reverse_discussion} can be expected to drive the model towards prioritising a certain subset of features in $p(\phi)$ at the expense of almost completely ignoring others; the objective function cannot penalise the model for making such a terrible error if the model never generates a configuration from that region.
This is well-documented in the literature \cite{Murphy2012,Minka2005}.
Recall that the flow-based approach to sampling is guaranteed to be ergodic; by construction, $\tilde{p}(\phi)$ has support on $\mathbb{R}^{|\Lambda|}$ (since it is bijectively related to a Gaussian distribution), which means there is a non-zero probability of \textit{any} flow model proposing \textit{any} configuration.
However, when the probability of sampling from important regions of configuration space is suppressed to the extent that they are almost never sampled from on the timescales of practical simulations, we might speak of an \textit{effective} breaking of ergodicity.

This appears to be worrying, and might remind the reader of the `mode collapse' problem faced when training generative adversarial networks.
Fortunately, in our case the problem is much less severe, because underestimates of $p(\phi)$ must necessarily be compensated by overestimates elsewhere in configuration space, which are then more likely to be penalised (this is a key strength of likelihood-based training).
This is not to say that the sort of dramatic error just described cannot occur, but in practice we found that it is far more likely to be caused by instability in the training due to using a large learning rate and small batch size, and not because the objective function is genuinely minimised by learning a density that is qualitatively different from the target due to model inflexibility.

\begin{figure}
    \includegraphics[width=.48\textwidth]{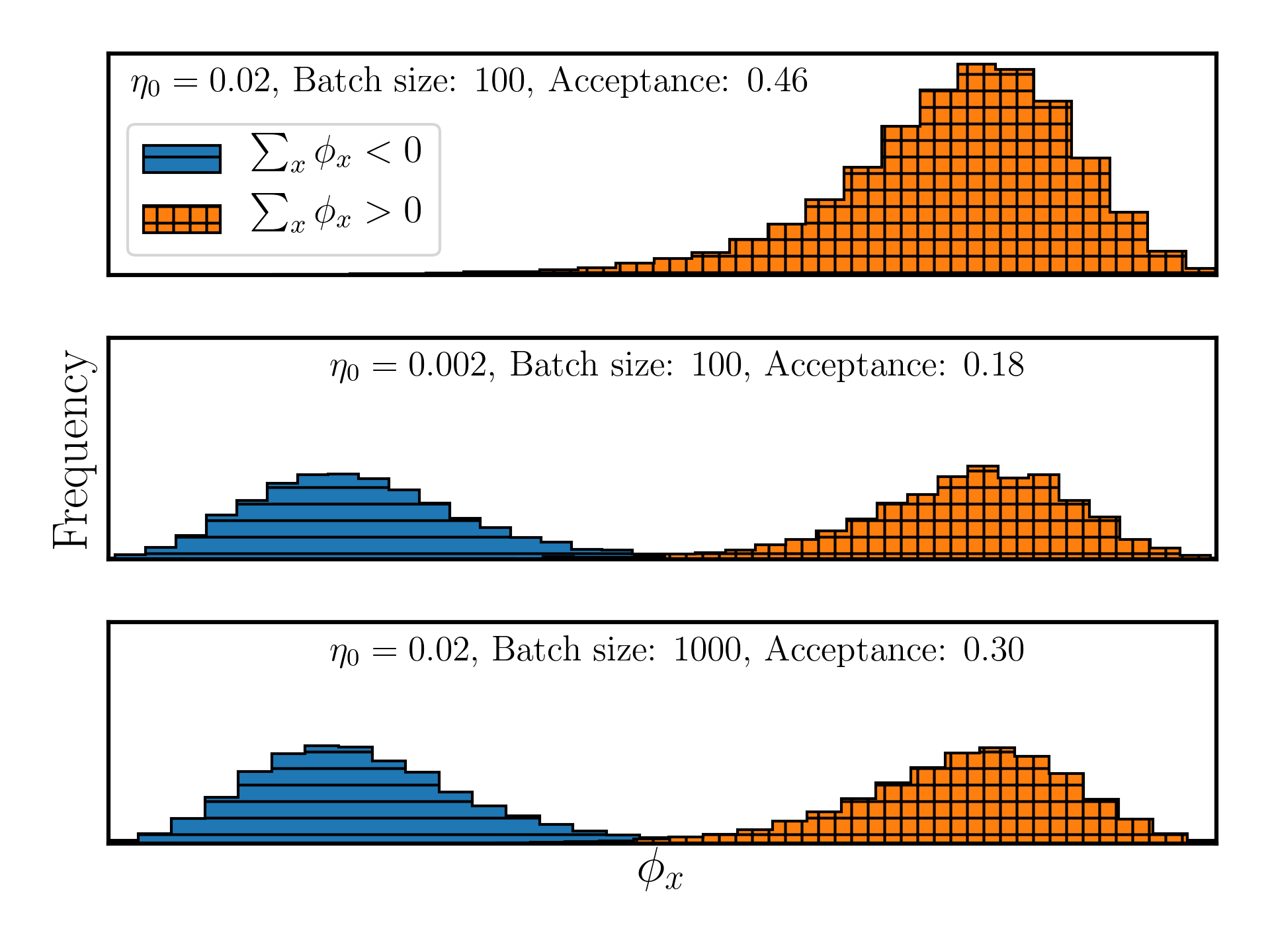}
    \caption{
    Histograms of field variables taken from three samples of $10^5$ configurations, generated by three different models. The colour labels the sign of the magnetisation. The $\phi^4$ parameters are $\{L, \beta, \lambda\} = \{6, 0.8, 0.5\}$. The models had two blocks of affine layers for which $\mathbb{Z}_2$-equivariance was not enforced. In the top sub-figure, the large learning rate and small batch size result in the breaking of the $\mathbb{Z}_2$ symmetry during optimisation. This is easily avoided by using more sensible learning rates and batch sizes. However, note that the acceptance rate for the top model is larger.
    }
    \label{fig:example_broken_z2}
\end{figure}

In the top sub-figure of Figure \ref{fig:example_broken_z2} we show an example of a model that has erroneously broken the $\mathbb{Z}_2$ symmetry by being trained too aggressively, the result being that it generates samples of field configurations whose magnetisations all possess the same sign.
The other two sub-figures demonstrate that the problem is easily resolved by decreasing the learning or increasing the batch size, thereby reducing the likelihood of making a sequence of optimisation steps which collapse one of the two modes.
Note that the `wrong' model attained a higher acceptance rate than both of the `correct' models.
However, this is misleading; the non-symmetric model is a highly \textit{inefficient} generator of proposals for the Metropolis-Hastings algorithm.
If the sampling was run for long enough, eventually a configuration from the collapsed mode would be generated, at which point the Markov chain would freeze due to the enormously suppressed probability of transitioning away.
Of course, this is exactly the level of inefficiency we should expect from a process that is attempting to perform a reweighting to a distribution with which there is very little overlap.

Although this example is rather contrived and very easily avoided by simply choosing sensible hyper-parameters, it serves as a warning not to rely on the acceptance rate as the sole indicator of model quality.
This conclusion was also reached by the authors of Reference~\cite{Hackett2021}.
Looking ahead to more complicated field theories, it will be sensible to check for violations of the symmetries which one knows to be present but which may be broken by the model.
In the $\phi^4$ case, we can easily see if the $\mathbb{Z}_2$ symmetry has been broken by looking at histograms of the field variables, and broken translational or rotational symmetries leave clear imprints in the correlation function.

\subsection{Critical slowing down of the training} \label{sec:discussion:csd}

Our results strongly suggest that the quality of the overall fit of our models to the target theory is limited by how extensively they have been optimised, not the inherent expressivity of the models.
For the $\phi^4$ couplings given in Table~\ref{tab:parameters} and our hybrid affine-spline models, Figure~\ref{fig:scaling} shows that these training costs scale in proportion to the number of degrees of freedom in the target density raised to a fairly high power --- in the region of 7--10.
An explanation for this disappointing result would be nice.

Of course, by moving to larger lattices we also increased the number of trainable parameters.
Since we fixed the neural networks to have a hidden layer of size $H = |\Lambda|$, the size of models grows as $|\theta| \propto |\Lambda|^2 = L^4$.\footnote{The weights of the connections between two neural network layers of widths $n$ and $m$ can be represented by an $n\times m$ matrix.}
The results of Figure \ref{fig:scaling} then state that, given a fixed target acceptance rate, the number of configurations $|\Phi_\text{train}|$ needed to train models scales with at least the square of the number of parameters requiring optimisation.
The numbers here are less important than the fact that we are \textit{not} seeing exploding training costs because the dimensionality of the optimisation problem is exploding (this was the point of looking for \textit{efficient} representations).

A convincing explanation for this apparent reduction in efficiency arises from a thorough analysis of the training procedure; specifically, the two terms in the gradient of the Kullbach-Leibler divergence described above.
\citeauthor{Huang2018b}~\cite{Huang2018b} showed both theoretically and empirically\footnote{Albeit in a slightly different context, in which the target density is given by a second generative model.} that, in several typical cases, the `expansion signal' due to the entropic term is feeble in comparison to the strength of the signal that drives the optimisation towards the mode(s) of the target, due to the term driving action minimisation.
Put differently, the probability of generating a configuration which produces a gradient, $\nabla_\theta \hat{D}_\text{KL}(\tilde{p} \; || \; p)$, that points away from the local mode of $p(\phi)$ is related to the spectrum of eigenvalues in the target's covariance matrix; the more \textit{ill-conditioned} this matrix (i.e. the more eigenvalues that are close to being zero, relative to the principal eigenvalue) the lower the probability.
The consequence of this is that the later stages of the training can be extremely inefficient, since a potentially tiny fraction of training inputs contribute to the the process of `expanding' around the mode(s) of the target, through which the model learns to match the low-density tails. 

We refer the reader to the original reference~\cite{Huang2018b} for the details of this argument and empirical results, although the argument is quite intuitive: an ill-conditioned covariance matrix arises when the probability density effectively resides on a low-dimensional sub-manifold --- this is sometimes referred to as having a low \textit{intrinsic dimension} with respect to the actual number of degrees of freedom in the distribution --- and in such situations we would expect the majority of training inputs generated by the model (which after all started life as an isotropic Gaussian) to fall outside this low-dimensional sub-manifold, producing a signal for the model to contract further towards the mode.
Meanwhile, the rate at which configurations are generated in a direction along which the model needs to expand is very low.
Our suspicion is, therefore, that the later stages of the training are predominantly spent compressing the model density onto a low-dimensional sub-manifold, and only very slowly expanding on this sub-manifold to better approximate the tails of the target.

\begin{figure}
        \includegraphics[width=.48\textwidth]{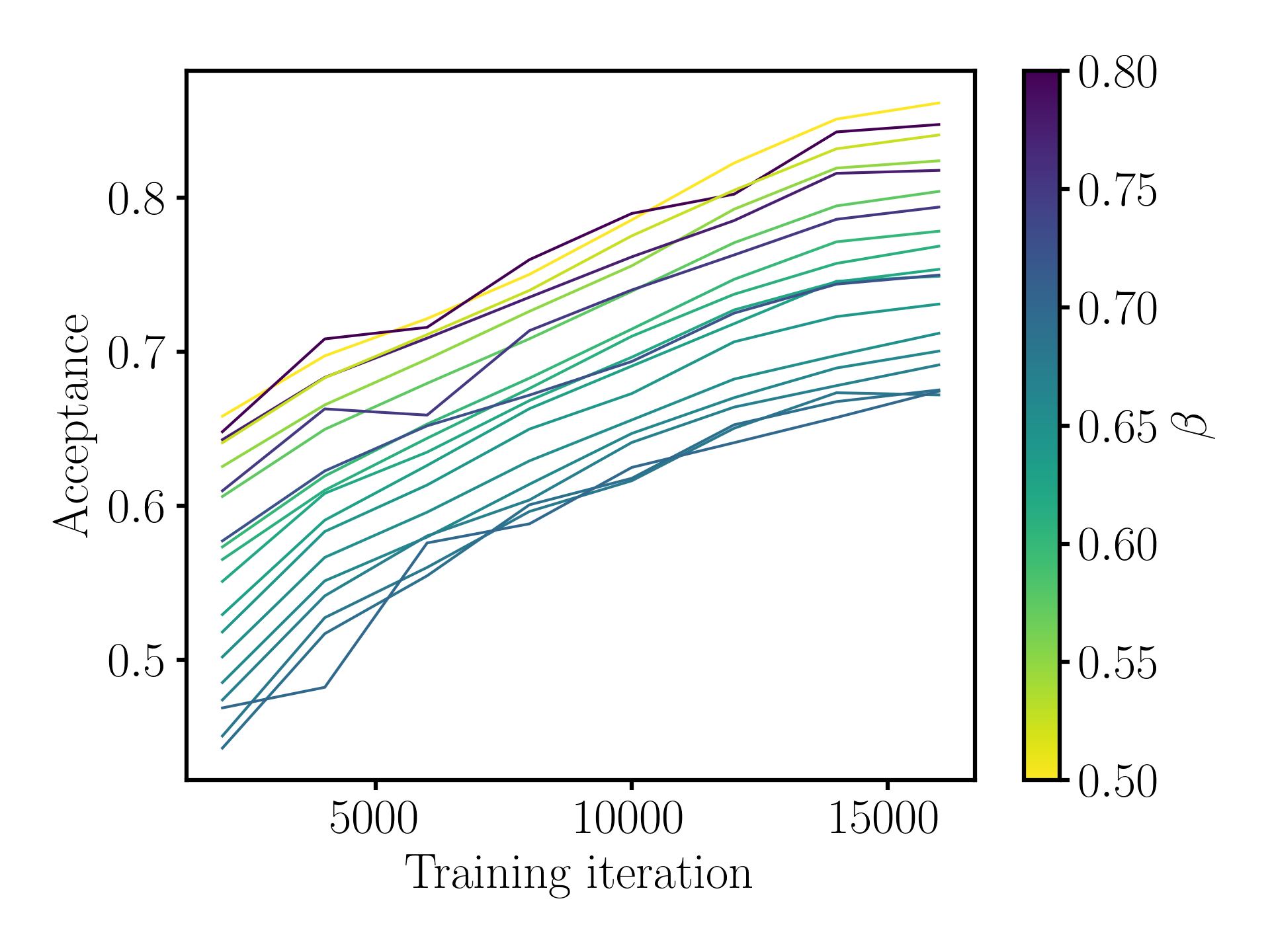}
    \caption{
    Metropolis-Hastings acceptance rate measured at intervals of 2000 iterations during the course of training the $L = 12$ models from Figure \ref{fig:beta_vs_acceptance}, in which $\lambda$ was held constant and $\beta$ varied so as to cross the phase transition.
    }
    \label{fig:acceptance_during_training}
\end{figure}

The observations of Reference \cite{Huang2018b} are highly relevant for data-driven applications because realistic data distributions tend to have low intrinsic dimension.
They are also relevant here; the action becomes increasingly ill-conditioned as the correlation length increases (i.e. effective mass tends to zero).
If further convincing is required, Reference \cite{MendesSantos2021} recently provided empirical evidence that, for a number of lattice models, critical points exist at minima of the intrinsic dimension.
Figure \ref{fig:acceptance_during_training} shows how the acceptance of the $L = 12$ models from Figure \ref{fig:beta_vs_acceptance}, with varying correlation length but a constant number of degrees of freedom, improved over the course of training.
Intriguingly, the profiles look very similar, other than being shifted with respect to each other.
Although interpreting this plot in light of the previous discussion is an exercise in speculation, it is possible that a consequence of increasing the correlation length is that a greater proportion of the target density is fit in the slow phase as it becomes more concentrated on a manifold of low dimension.

To be clear, it is unsurprising that model optimisation should becoming increasingly challenging as a critical point is approached.
What is perhaps more interesting is the way in which a long correlation length afflicts the particular training scheme used here, although it is very possible that effect just described is subdominant on the small lattices studied in this work (in which case we need to identify the dominant source of inefficiency).
Much more work will be needed to disentangle the various factors contributing to the overall scaling of training costs and paint a more quantitative picture, but if this alternate manifestation of critical slowing down turns out to be a key bottleneck then attention should turn to establishing whether it can be tamed more readily than the familiar version that hampers traditional MCMC.
In Section \ref{sec:outlook:scalability_physics} we discuss some possible avenues for further investigation.


\section{Conclusions and outlook} \label{sec:outlook}

We have verified the key results of Reference \cite{Albergo2019}: firstly, that there exist approximately trivializing maps in two-dimensional lattice $\phi^4$ theory that are accessible \textit{in practice} to normalizing flow models with tractable Jacobian determinant; and secondly, that using such models as generators of proposals for a Metropolis-Hastings simulation represents a complete transfer of the computational costs normally associated with critical slowing down to the cost of training the flow model.
As it often the case in Machine Learning, the efficiency of the procedure depends on the tuning of hyperparameters, which we have investigated in this work. We have shown that fairly modest modifications of the original prescription --- inserting a more expressive transformation at the final layer of the flow and drastically reducing the size of the neural networks --- lead to much more efficient representations of approximately trivializing maps for this system.
The extent of the associated reduction in training costs is such that the systems studied here and in Reference~\cite{Albergo2019} are accessible with the sorts of computing resources typically found on personal computers.
However, our main finding is that the rate at which training costs scales as we move towards the continuum limit is extremely large, increasing far more quickly than the size of the models.
Our work demonstrates a rather urgent need to understand and mitigate inefficiencies in the training algorithm itself when the target of optimisation is a lattice field theory in the critical regime, and this must be done in parallel with efforts towards building more sophisticated flow models.

Below, we outline our immediate intentions for further research and describe several potential options for improving on the scaling of training costs.

\subsection{Non-linear $\sigma$ models} \label{sec:outlook:sigma}

Further work on our part will focus on the $\mathrm{O}(N)$ and $\mathrm{CP}^{N-1}$ non-linear $\sigma$ models.
The latter class of models are a good milestone on the journey to QCD since they exhibit many of the interesting non-perturbative features observed in QCD whilst being much more amenable to numerical study, being two-dimensional theories without even a fundamental gauge field.
There is precedent for expecting $\mathrm{CP}^{N-1}$ to challenge this method; previous studies uncovered pathological critical slowing down of the topological charge $Q$, consistent with $\tau_Q \sim e^{c\xi}$ \cite{DelDebbio2004}, and it was precisely this situation which the original formulation of trivializing maps failed to resolve \cite{Engel2011}.

Non-linear $\sigma$ models also have a non-trivial field space, though one can parameterise any $\mathrm{O}(N)$ element or $\mathrm{CP}^{N-1}$ representative using the unit $n$-spheres.
Fortunately for us, several recent efforts have augmented the set of tools at our disposal with several that are dedicated to problems involving angular variables \cite{Gemici2016,Rezende2020,SohlDickstein2020,Passenheim2020}.
The generalisation is straightforward in principle; as well as being invertible and continuously differentiable the flow is required to respect the topological properties of the field space.
However, one should be aware that in most implementations of normalizing flows, the present work included, a Euclidean setting is implicit in the change-of-variables formula defined by Equation~\eqref{eq:ch_var_formula} as well as, more fundamentally, in the neural networks, which naturally act on Euclidean vectors.

\subsection{Improving on the scalability with physics} \label{sec:outlook:scalability_physics}

More work is required to determine just how large a role this analogue of critical slowing down (see Section \ref{sec:discussion:csd}) plays in the exploding cost of training.
Our suspicion is that the correlation length of the target theory will contribute significantly as one moves to larger lattices.
If our understanding is correct, it may be more efficient to learn a trivializing map to the corresponding free theory, implying that (with a suitably chosen bare mass) the latent density $r(z)$ resides on a similarly low-dimensional manifold as the target.
We can easily generate real-space configurations of non-interacting fields by simply rescaling and performing a Fourier transform, and the log-density term is simply given by the action of the free theory.
A cursory look at some models trained with free fields acting as latent variables indicates that starting from free fields does not dramatically alter the overall quality of optimisation if the target theory is strongly interacting.
However, this could be because any structure in the latent density is obfuscated during the fast zero-forcing stage, which may fail to keep the intrinsic dimension low.
A more robust strategy might involve gradually flowing through the space of couplings from the free theory to the theory of interest, more akin to the original trivializing maps of Reference~\cite{Luscher2009}.
This could be achieved by updating the couplings during training,\footnote{There is no reason why the couplings of the theory have to remain constant during training. Rather, they can be thought of as another tunable parameter of the optimisation, alongside the batch size and learning rate~\cite{Hackett2021}.} or by a layer-wise approach to training that forces the trivializing map to follow a constrained path through the space of couplings.

Notwithstanding issues related to specific training schemes, there are reasons to be a touch more optimistic about the scalability of this method.
Most actions or Hamiltonians that are of interest in physics contain exclusively local interaction terms, possess symmetries under certain transformations, and are often approximately or exactly self-similar over multiple scales.
Each of these qualities might be exploited to build highly constrained models that yield efficient representations of trivializing maps.

Locality is partly responsible for the success of convolutional architectures, which pass localised filters (convolution kernels) over the data.
`Features' can be extracted at multiple scales by downsampling the data between convolutional layers, a procedure which shares much in common with Kadanoff's block decimation \cite{Kadanoff1966}, the precursor to renormalization group transformations \cite{Wilson1973}.
One of the implications of renormalizibility is that coarse-grained representations of the fields contain useful information encoded as relevant variables, providing a physical motivation for the use of multi-scale architectures which progressively resolve features at finer scales.
Multi-scale flows form the basis of the information-preserving renormalization group algorithm of \cite{Li2018}, and similar architectures have performed well in the classic application of image synthesis \cite{Dinh2016}, where the same concepts of locality and coarse-grained descriptions hold true, albeit less formally.
We also note that convolutional networks that are equivariant or covariant with respect to other types of symmetry transformations have been introduced in References \cite{Cohen2016,Cohen2018,Cohen2019,Bekkers2019,Tomiya2021}.

\subsection{Alternatives for improving scalability} \label{sec:outlook:scalability_nonphysics}

One might describe the ideas suggested above as `physics-based'; the ambition is to somehow incorporate known physics to construct models and training schemes that are `better informed' about the nature of the target density.
However, there are a number of `technical' options which might (at least partially) circumvent the issue with training.
One of these is to devote some effort to generating a training set of configurations via traditional MCMC (potentially as a bootstrapping of the generative training update), and include these in the optimisation.
Even a relatively small number of training examples sampled from the low-density tails may help to improve the rate at which the model expands to fit them, though we note that Reference~\cite{Albergo2019} did not report any significant improvements by using a pre-generated training set.
There are also ways forward in which we just accept that sampling from regions of low density is a weakness of the generative approach.
For example, it would be trivial to incorporate regular local updates alongside generative proposals in the sampling phase, which may to get the best of both worlds; local moves could offset biases in regions of low target density more efficiently than the Metropolis test alone, which as we have seen results in long periods of consecutive rejections.
Another possibility, more appealing in terms of scalability, is to embed generative proposals within a multi-level sampling algorithm such as those developed in References \cite{Luscher2001,Meyer2002}.
In effect, this means training models to trivialize sub-volumes of the lattice in which the correlation length is cut off by $L_\text{subvol}$, trading improved efficiency in the training stage for the additional cost of having to stitch together these sub-volumes.

Even in very optimistic scenarios where improvements to the training scheme alleviate the worst effects of critical slowing down, we would still expect larger batches and longer training runs to universally lead to improvements (up to a point where the expressivity of the model prevents it from being able to fit the target any better), since there is no over-fitting to speak of.
Distributing batches over multiple processing units, a.k.a `data-parallel' training, is therefore likely be an essential ingredient of future attempts to scale this method up to larger systems.
The efficiency of the data-parallel approach is reduced as node memory limits are exceeded and the batch size per node must be decreased.
Hence, it bodes well that we found shallow neural networks to outperform their deep counterparts, suggesting that memory costs will increase relatively slowly.
To contrast this point, several state-of-the-art models based on residual networks with hundreds or thousands of layers have such high memory requirements that \textit{the layers themselves} must be distributed over multiple nodes.
In fact the situation as regards memory is even better since normalizing flows lend themselves to low memory requirements by their very design; backpropagation through reversible layers (such as coupling layers) can be performed without storing the intermediate vectors \cite{Gomez2017}, meaning that memory requirements do not increase in proportion to the number of coupling layers.

\subsection{Wider context} \label{sec:outlook:context}

On a final note, we find it thought-provoking that several generic features of `real-world' data, such as locality, symmetries and scale invariance, arise in lattice field theories, but in a manner that is more precisely defined, e.g. by terms in the action, renomalization group transformations etc.
Meteorological data, for want of an example, contains correlations over multiple scales and emergent phenomena arising from purely local interactions.
We also draw attention to an intriguing study in which machine learning techniques were used to \textit{measure} the presence of various symmetries (e.g. $\mathbb{Z}_2$, $\mathrm{O}(2)$) in pieces of artwork~\cite{Barenboim2021}.
The framework of lattice field theory provides low-level control over these quasi-universal properties through our ability to simply write down an action.
Hence, lattice field theory is in many ways an ideal test-bed for improving our understanding of how these properties can be efficiently encoded into statistical models, even when the technique or model under study is destined for completely unrelated applications.
A further convenience of working with lattice field theories is that there are several options for generating reproducible data for training or validation in a way that can be done on-demand (with the caveat that state-of-the-art ensembles are extremely expensive to generate) instead of requiring permanent storage.
Perhaps it is not too outrageous to imagine that a set of lattice field theories may, in future, become a standard suite for testing and benchmarking new models and algorithms.

\begin{center}
\textbf{Acknowledgements}
\end{center}

LDD is supported by an STFC Consolidated Grant, ST/P0000630/1, and a Royal
Society Wolfson Research Merit Award, WM140078.
JMR is supported by STFC, grant ST/T506060/1.
MW is supported by STFC, grant ST/R504737/1.
This work has made use of the resources provided by the Edinburgh Compute and Data Facility (ECDF) \cite{ECDF}.


\bibliography{bibliography}


\appendix

\section{$\phi^4$ theory on the lattice} \label{sec:appen:phi_four}

The standard $\phi^4$ action in two-dimensional Euclidean space is
\begin{align}
    S[\varphi] = \int d^2 x \bigg[
        \frac{1}{2} \big( \partial_\mu \varphi(x) \big) \big( \partial_\mu \varphi(x) \big) 
    + \frac{1}{2} m_0^2 \varphi(x)^2 \nonumber\\
    + \frac{1}{4!} g_0 \varphi(x)^4 \bigg] \, ,
\end{align}
where $m_0$ is the bare mass and $g_0$ is the bare coupling for the quartic interaction term.

We can define a discretised analogue of this theory on a periodic lattice $\Lambda$ with lattice spacing $a$ and spatial extent $L_\mu = a N_\mu$ using the following steps:

\begin{enumerate}
    \item Use the vanishing boundary term (due to periodicity) to replace the derivative term with the Laplacian,
    \begin{equation}
        \big( \partial_\mu \varphi(x) \big) \big( \partial_\mu \varphi(x) \big)
        \rightarrow - \varphi(x) \partial^2 \varphi(x) \, .
    \end{equation}
    \item Adopt the following discretised version of the Laplacian:
        \begin{align} \label{eq:phi_four:laplacian}
        \partial^2 \varphi_x &\rightarrow \delta^2 \varphi_x \nonumber\\
        &= \frac{1}{a^2} \sum_{\mu=1}^2 ( \varphi_{x+a e_\mu} + \varphi_{x-a e_\mu} - 2\varphi_x) \, ,
        \end{align}
        where $e_\mu$ represents a unit vector in the $\mu$-th dimension.
    \item Replace the integral with a sum,
        \begin{equation}
            \int d^2 x \rightarrow a^2 \sum_{x\in\Lambda} \, .
        \end{equation}
    \item For convenience, define dimensionless couplings $m_0^2 \to m_0^2 a^2$ and $g_0 \to g_0 a^2$.
    
\end{enumerate}
This leads to the following lattice action:
\begin{equation} \label{eq:phi_four:standard_lattice_action_with_laplacian}
    S(\varphi) = \sum_{x\in\Lambda} \bigg[
        \frac{1}{2} \varphi_x ( -\delta^2 + m_0^2 ) \varphi_x + \frac{g_0}{4!} \varphi_x^4 \bigg] \, .
\end{equation}
Using \eqref{eq:phi_four:laplacian} and the translational invariance of the action yields
\begin{equation} \label{eq:phi_four:standard_lattice_action}
    S(\varphi) = \sum_{x\in\Lambda} \bigg[
        - \sum_{\mu=1}^2 \varphi_x \varphi_{x+e_\mu} 
        + (2 + \frac{m_0^2}{2}) \varphi_x^2
        + \frac{g_0}{4!} \varphi_x^4
    \bigg] \, .
\end{equation}

Equation \eqref{eq:phi_four:action} is related to \eqref{eq:phi_four:standard_lattice_action} through
\begin{equation}
    \varphi = \sqrt{\beta}\phi, \qquad 2 + \frac{m_0^2}{2} = \frac{1 - 2\lambda}{\beta}, \qquad \frac{g_0}{4!} = \frac{\lambda}{\beta^2} \, .
\end{equation}

\section{Estimation of integrated autocorrelation time} \label{sec:appen:autocorr}

\begin{table*}
    \centering
    \begin{ruledtabular}
    \begin{tabular}{c c c c c c c c c}
        $L$ & \# hidden elements & \# affine blocks & \# spline segments & batch size & \# training iterations & $\eta_0$ & time & time (s) \\
        \hline
        6 & 36 & 2 & 8 & 300 & 350 & 0.02 & 9s & $9 \times 10^0$ \\
        8 & 64 & 2 & 8 & 500 & 750 & 0.01 & 37s & $3.7 \times 10^1$ \\
        10 & 100 & 3 & 8 & 1250 & 2000 & 0.005 & 5.0m & $3.0 \times 10^2$ \\
        12 & 144 & 3 & 8 & 4000 & 5000 & 0.003 & 43m & $2.6 \times 10^3$ \\
        14 & 196 & 3 & 8 & 7000 & 14000 & 0.001 & 4.5h & $1.6 \times 10^4$
    \end{tabular}
    \end{ruledtabular}
    \caption{
    Measurements of the real time taken to train our models to reach an acceptance rate of 70\% for the systems studied in Reference \cite{Albergo2019}.
    The models were trained on a desktop PC with an Intel i7-7700K quad-core CPU and 16GB RAM.
    Since we were aiming for speed of training rather than reaching the highest possible acceptance rates, we increased the initial learning rate $\eta_0$ with respect to our main study, although we do not recommend doing this in general.
    }
    \label{tab:paper:albergo70}
\end{table*}

In practice, the integrated autocorrelation time defined by Equation \eqref{eq:tau_int} must be estimated from a Markov chain of finite length $N$.
However, the statistical error on the autocovariance estimator,
\begin{equation}
\hat{\Gamma}_\obs(t) = \frac{N}{N-1} \bigg[ \frac{1}{N} \sum_{n=1}^{N-t} \obs(\phi^{(n+t)}) \obs(\phi^{(n)})
- \bar\obs^2 \bigg] \, ,
\end{equation}
increases with $t$, so it is preferable to truncate the sum at some separation $W < N$.

We thus have the estimator
\begin{equation} \label{eq:tau_int_estimator}
    \hat{\tau}_\obs(W) = \frac{1}{2} + \sum_{t=1}^W \frac{\hat\Gamma_\obs(t)}{\hat\Gamma_\obs(0)} \, ,
\end{equation}
and must attempt to find the value of $W$ which minimises the sum of:
\begin{enumerate}
    \item The bias due to truncating the sum,
        \begin{align} \label{eq:trunc_bias}
            \varepsilon_\text{trunc}(W) &\equiv
            \text{bias}\big[ \hat{\tau}_\obs(W)\big]  \nonumber \\
            &= -\sum_{t=W+1}^\infty \frac{\Gamma_\obs(t)}{\Gamma_\obs(0)}
            \approx -\tau_\obs e^{-W/T_\obs} \, ,
        \end{align}
        where we have assumed that $W$ is sufficiently large that the autocorrelation takes a pure exponential form, with $T_\obs$ being the characteristic relaxation time of the slowest mode of $\obs$.
    
    \item The statistical error approximated by the Madras-Sokal formula \cite{Madras1988,Sokal1997},
        \begin{equation}
            \varepsilon^2_\text{stat}(W)\equiv
            \var \big[ \hat{\tau}_\obs(W) \big] \approx \frac{2(2W + 1)}{N} \tau_\obs^2 \, ,
        \end{equation}
        which uses the approximation $T_\obs \ll W \ll N$.
\end{enumerate}
We would therefore like to find the minimum of $\Delta(W) = |\varepsilon_\text{trunc}(W)| + |\varepsilon_\text{stat}(W)|$.
To do so we follow the `automatic windowing' procedure detailed in Reference \cite{Wolff2003} (Sec. 3.3).

First, note that we can re-cast the integrated autocorrelation time in terms of the equivalent pure exponential decay,
\begin{equation}
    \tau^*_\obs = - \bigg[ \log \bigg( \frac{2 \tau_\obs - 1}{2 \tau_\obs + 1} \bigg) \bigg]^{-1} \, ,
\end{equation}
which is equal to zero, rather than $1/2$, for uncorrelated data, and so offers improved precision in situations where decorrelation occurs very quickly.

We furthermore assume that it is valid to substitute the slowest mode $T_\obs$ for $\lambda \tau^*_\obs$ with $\lambda$ being a small constant factor, to be tuned such that smallest value of $W$ for which
\begin{equation}
    \lambda \frac{\partial \hat{\Delta}(W)}{\partial W} = -e^{-W/(\lambda\hat{\tau}^*_\obs)} + \frac{\lambda \hat{\tau}^*_\obs}{\sqrt{WN}}
\end{equation}
drops below zero occurs, generally, at a point at which $\hat{\tau}_{\text{int},\obs}(W)$ levels off to a plateau.

Since we generally encountered very small integrated autocorrelation times, the approximation in Equation~\eqref{eq:trunc_bias} is probably a poor one.
By the same token, however, statistical errors were minimal.
Ultimately, a little tuning of $\lambda$ by visual inspection of $\hat{\tau}_\obs(W)$ for a small number of experiments was sufficient.

\vspace{.35cm}
\section{Comparison with literature results} \label{sec:appen:albergo}

\citeauthor{Albergo2019} \cite{Albergo2019} used flows comprising solely affine coupling layers parameterised by fully-connected networks to generate $\phi^4$ configurations on lattices ranging from $6^2$ to $14^2$ sites.
They reported training times of 1-2 GPU-weeks in order to reach an average acceptance rate of 70\% in the Metropolis-Hastings phase.

Although theirs was a proof-of-principle study, we were nonetheless curious to check how quickly our hybrid affine/spline models could be trained to reach a 70\% acceptance rate, using the same action and couplings provided in Reference \cite{Albergo2019} (though our previous results for $|\Lambda| = 6^2 - 14^2$, $\lambda=1/2$ and $L/\xi = 4$ correspond to essentially the same systems).
We trained these models on a desktop PC.
The real time taken to train models to reach at least 70\% acceptance are given in Table \ref{tab:paper:albergo70}.

Presumably, the main reason that the times reported in the original study are so much larger than what we find is the use of much larger neural networks; for the $14^2$ lattice the original study used networks with width 1024 and depth 6 whereas for the same system our models used comparatively minuscule networks of width 196 and depth 2 (i.e. a single hidden layer).
Still, given the near order-of-magnitude increase in real train times for each lattice size increase, with this setup and a target of 70\% acceptance we would likely encounter times measured in weeks as soon as $L = 18$ or 20.
Despite the modest hardware, it is abundantly clear that scaling this technique up to system sizes at which critical slowing down becomes a serious problem will require significantly more than fine-tuning the current approach.

\end{document}